\documentclass{aastex631}
\usepackage{comment}
\usepackage{newtxtext,newtxmath}

\begin{document}

\title{Stellar flare study of nearby young moving group members with
TESS Data}

\author{Andrew Tran}
\email{aat30279@uga.edu}
\affiliation{Department of Physics and Astronomy, University of Georgia \\
Sanford Dr. Athens, GA, 30606, USA}

\author[0000-0002-5815-7372]{Inseok Song}
\email{song@uga.edu}
\affiliation{Department of Physics and Astronomy, University of Georgia \\
Sanford Dr. Athens, GA, 30606, USA}



\begin{abstract}

We analyze TESS data to explore stellar flares and rotational characteristics in members of Nearby Young Moving Groups (NYMGs). Our study focuses on 417 members of NYMGs aged 10-150 Myr. Using detrended light curves from the TESS Science Office Quick-Look Pipeline, coupled with our own additional detrending scheme for fast rotators, we systematically detect and characterize 6,288 stellar flares from 27,416 flare candidates. We analyzed light curves from Cycles 1-4 of the TESS mission, finding that for each NYMG member analyzed, at least one stellar flare was present. Flare candidates are initially detected using the AltaiPony flare package, followed by a recovery flare amplitudes, durations, and local continuum background levels. We examine the relationship between flare energy, age, and mass, finding a reduced flaring rate for late-type stars with age for high energy flares, as well as 5.5 times more flares detected in the 10-minute cadence TESS data compared to 30-minute cadence data. Additionally, flare events with extreme energies ($E \geq 10^{34}$ erg) on M-dwarf and solar-type stars, providing implications for further exploration into exoplanet habitability.

\end{abstract}



\section{Introduction} \label{sec:intro}

Nearby young moving groups (NYMGs) are gravitationally unbound groups of stars with the same relative motion in the age range of 10 Myr to 200 Myr and within $\approx$ 100 pc from the Sun. These groups are critical for studying the early stages of stellar and planetary evolution due to their youth ($\leq$200 Myr) and proximity ($\approx$100 pc), which allow for detailed observations and analyses of stellar formation processes, disk evolution, and planet formation dynamics \citep{Torres2008, Zuckerman2011}. NYMG studies have provided valuable insights into stellar evolution in these associations \citep{2004ARA&A..42..685Z} and have contributed to more reliable age-dating methods,  \citep{Lee_2022, BarradoNavascués_1999, Gagne2017}. Additionally, NYMG members, such as those in the $\beta$ Pictoris Moving Group, have historically been prime candidates for studying planetary system formation \citep{2001ApJ...562L..87Z}. These groups are valuable for direct imaging and spectroscopy of young planets, offering insights into planet formation processes that are inconceivable currently in older systems \citep{Marley2007, Macintosh2015}. Furthermore, their homogeneous environment facilitates accurate determinations of stellar properties, such as chemical abundances, which are essential for contextualizing any exoplanets discovered \citep{Spiegel2012}. NYMGs also provide a near-complete census of low-mass M-dwarf stars, which are of significant interest for habitable planet studies \citep{Dressing2013, Henry2006}.

Stellar flares, which are sudden releases of energy due to magnetic reconnection events, are crucial in understanding the magnetic, rotational, and exoplanet-related characteristics (such as the effect of flares on the atmosphere's chemical composition) of stars \citep{2019ApJ...871..241D}. Flares are particularly important for investigating the conditions necessary for planetary habitability, as they influence the atmospheric chemistry and magnetic environment of exoplanets \citep{Hazra_2021}. Previous studies utilizing photometric data from surveys like Kepler \citep{2019ApJ...871..241D} and the Transiting Exoplanet Survey Satellite (TESS) \citep{Ricker_2014} have provided a wealth of information on flare activity in young stars. Recent works have focused on characterizing flares across various spectral types and ages to better understand the magnetic environments in which planets form \citep{Feinstein_2020}. Additionally, coronal mass ejections and flares are thought to have significant impacts on Jovian-like exoplanets \citep{Howard_2021}, and flare frequency distributions (FFDs) have been used as a tool to investigate coronal heating mechanisms \citep{Verbeeck_2019}.

While previous studies have laid the groundwork for understanding flare activity, particularly in M-dwarfs and other low-mass stars, gaps remain in comprehensively characterizing flares in younger populations. For instance, the flaring rate reaches its peak for M5-M7 stars but drastically diminishes for cooler M-dwarfs \citep{Murray_2022}. This decline is thought to reduce the potential for photosynthetic processes on surrounding exoplanets \citep{Feinstein_2020}. \cite{Caldiroli_2025} similarly found that M-dwarf flares have a minimal effect on the atmospheric retention of Neptune-size planets, including ones in the habitable zones of their stars. In G-type stars, studies have found that superflares—events with energies larger than $10^{33}$ ergs—occur not only on fast rotators but also on slow rotators with periods comparable to that of the Sun \citep{2022ApJ...941..193C}, corroborating earlier findings \citep{Maehara2012}. Additionally, flare energies and durations in M-dwarfs have been found to follow scaling laws similar to those of solar white light flares \citep{Paudel2024}, suggesting common underlying physical processes.

In this work, we offer additional insights over previous flare investigations. First, while recent work has primarily focused on older stellar populations, our study centers on stars in NYMGs within the 10-150 Myr age range, using the membership list from \cite{LeeSong2019}, and utilizing 30 minute and 10 minute cadence TESS minute data.

We apply comprehensive flare detection procedures across light curves from members of nine NYMGs, spanning multiple observation sectors from four TESS cycles (2018-2023). By categorizing our sample by moving group (age) and spectral type, we investigate cumulative flare frequency distributions (FFDs) and as well as relationships between flaring activity and various astrophysical parameters. Our analysis provides further sampling and insights into the behavior and characteristics of stellar populations in the 10-150 Myr age range, adding valuable data to the ongoing study of stellar and planetary evolution in NYMGs.

\section{Data and Methodology} \label{sec:style}

\begin{figure}
    \centering
    \includegraphics[width=0.8\linewidth]{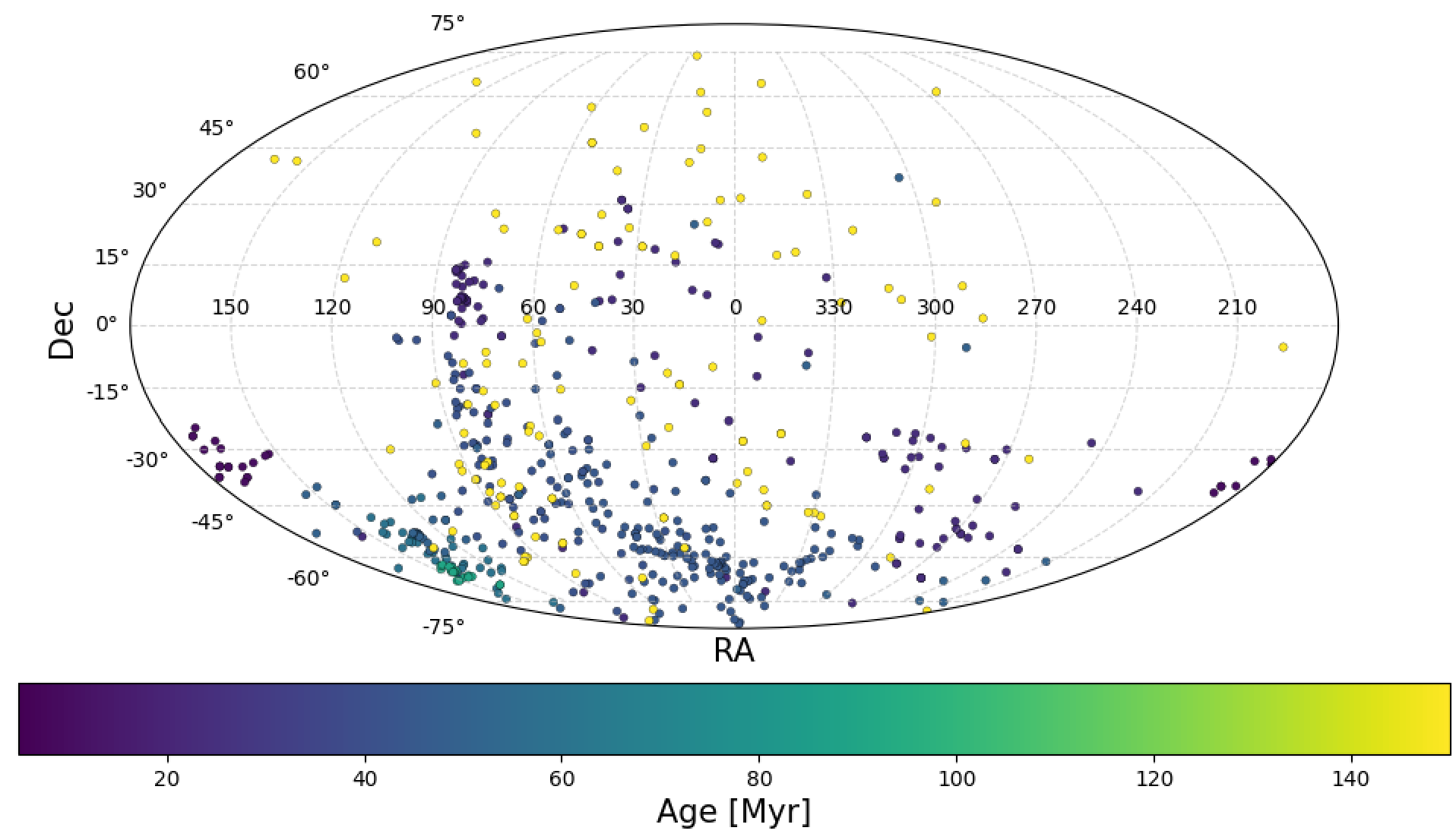}
    \caption{Distribution of the selected NYMG sample in the sky that had QLP data available for download, color coded by age of the associated moving group.}
    \label{fig:ra_dec_projection}
\end{figure}

\subsection{TESS Data}
We utilized data from the first four observation cycles (2018-2023) of the Transiting Exoplanet Survey Satellite (TESS), where Cycles 1-2 correspond to 30-minute cadence data, and Cycles 3-4 correspond to 10-minute cadence data. The distribution of the NYMG members with available TESS lightcurves is shown in Figure \ref{fig:ra_dec_projection}, with the majority of the members concentrated in southern declinations, due to the distribution of NYMGs. By cross-matching the list of NYMGs \citep{LeeSong2019} in MAST for matching TESS data, we initially identified 2,240 TESS light curves and downloaded them, and removed contaminated cases, using the following relation from \cite{Rebull_2022}:

\vspace{1.5mm}
\begin{equation}
    {\: \: \: \: \: \: \: \: \: \: \: \: \: \: \: \: \text{log(mean LC flux)} = -0.358 \times G + 8.45}
\end{equation}
\vspace{1.5mm}

\begin{figure}
    \centering
    \includegraphics[scale = 0.75]{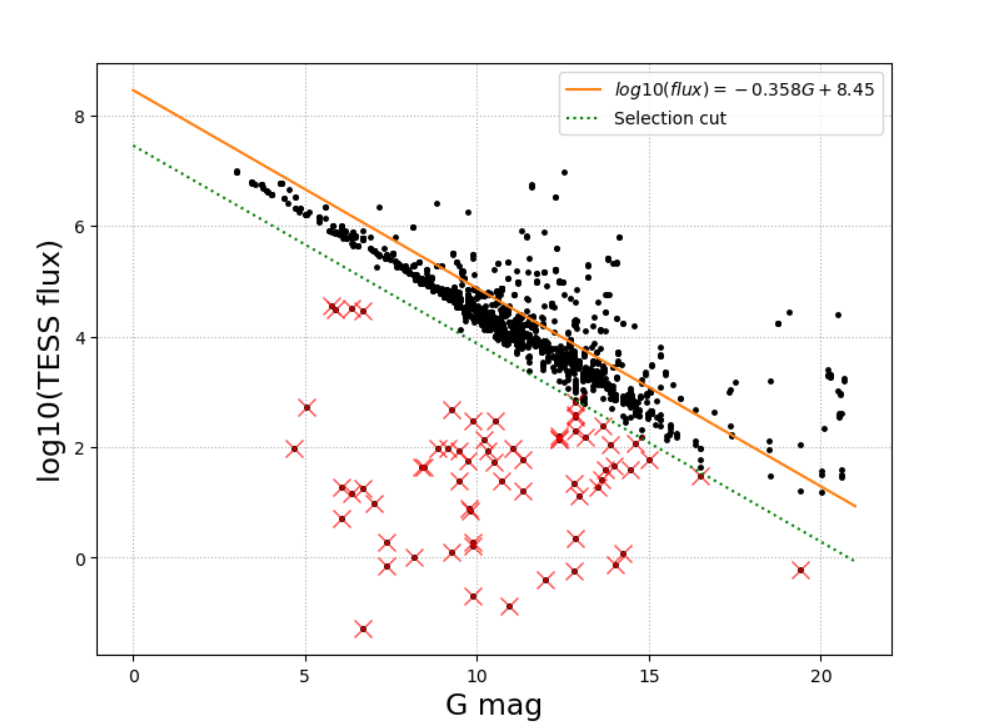}
    \caption{The contamination plot for the TESS data in this study. Data points lying above the orange line are potential contaminants, as they have additional stars in the TESS aperture so that for a given G-band magnitude, we see higher flux values. Data points marked with a red 'X' potentially contain interference from non-astrophysical sources.}
    \label{fig:1}
\end{figure}

This relation provides a predictive model for estimating TESS instrumental flux (in units of e$^{-}$ sec$^{-1}$) from \textbf{Gaia DR2}'s G-band magnitude. The results from applying this relation to our sample are displayed in Figure \ref{fig:1}. In this figure, points lying above the orange curve are inconsistent with what would be expected from their G-magnitudes, implying TESS fluxes were affected from background stars \citep{Rebull_2022}. The green dashed line represents the selection cut used to remove likely contaminated apertures. It was defined as an offset of 0.65 dex below the best-fit flux–magnitude relation to exclude targets with excess TESS flux relative to their G-band magnitude. Additionally, stars below this cut recorded too low TESS fluxes for given Gaia brightness indicating unreliable TESS measurements for these stars. Upon closer examinations, TESS photometric data for these sources were affected by events such as the bright diffraction spikes from adajacent bright stars, etc. We removed stars above the orange line or below the green line. In total, we removed 378 contaminated light curves.



\subsection{Light Curves}

\begin{figure*}
    \centering
    \includegraphics[width = 0.9\textwidth, height = 10 cm]{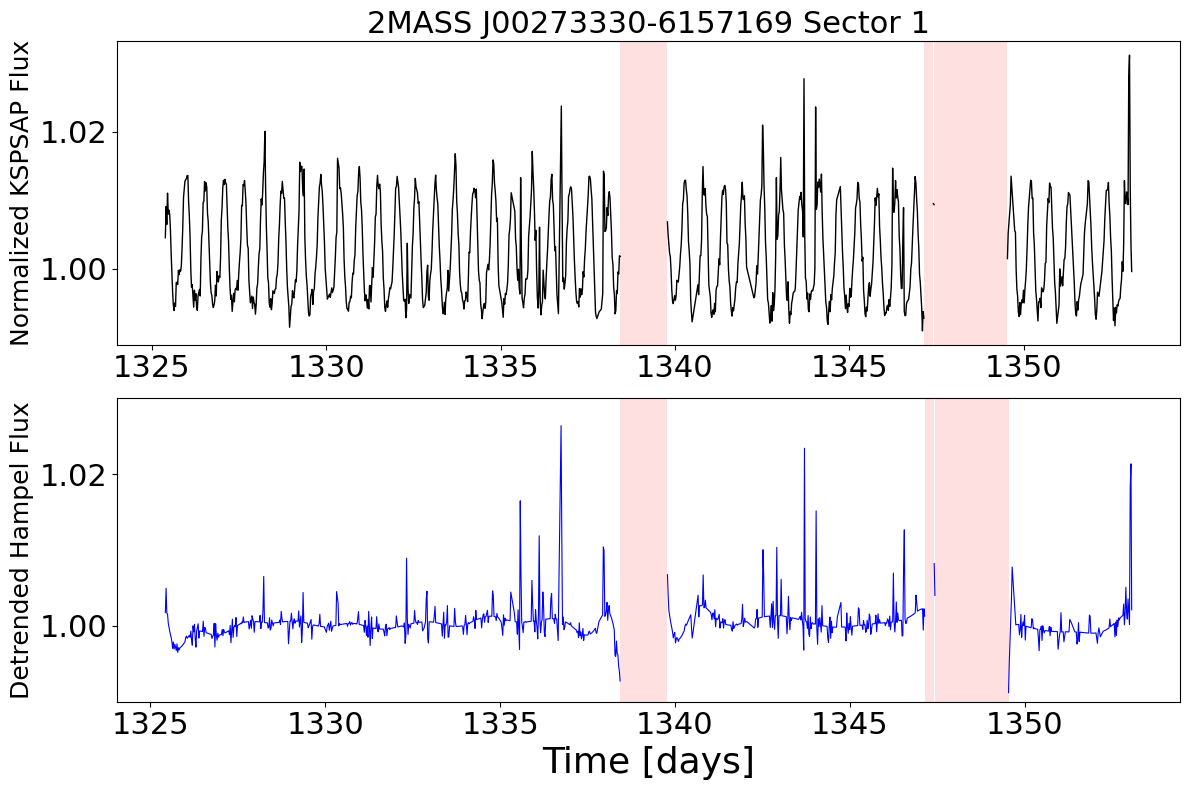}
    \caption{Top: KSPSAP detrended flux of 2MASS J00273330-6157169. A strong rotational modulation of around $0.55$ days (calculated using Lomb-Scargle) is still present even after detrending. Bottom: the KSPSAP Flux with the Hampel filter applied. In both plots, the red shaded regions indicate data gaps.}

    \label{fig:hampel}
\end{figure*}

We used light curves from the TESS Science Center’s Quick-Look Pipeline (QLP). We were able to cross-match and retrieve light curves for 417 stars between our membership list and the available QLP sample. In general, most of the raw light curve exhibited a variable, sinusoidal-like rotation-modulated variability with respect to time, allowing us to recover a rotational period as well. However, many light curves (LCs) detrended by the QLP pipeline were insufficiently detrended or possibly not detrended at all, particularly for fast rotators. This issue could have affected the flare identification process. To remove the remaining rotational signals in QLP LCs, we applied a Hampel-based rolling median filter to all light curves with estimated rotational periods less than $~0.6$ days, obtained from each star's Lomb-Scargle Periodogram, to remove the stellar baseline, then used median absolute deviation (MAD) outlier detection to suppress impulsive outliers, before normalizing the light curve using the rolling median. This removed long-term trends while preserving flare amplitudes. A comparison between the performance of the QLP pipeline versus the Hampel filter is shown in Figure \ref{fig:hampel}.

\subsection{Flare Detection}

\begin{figure}
    \centering
    \includegraphics[scale = 0.75]{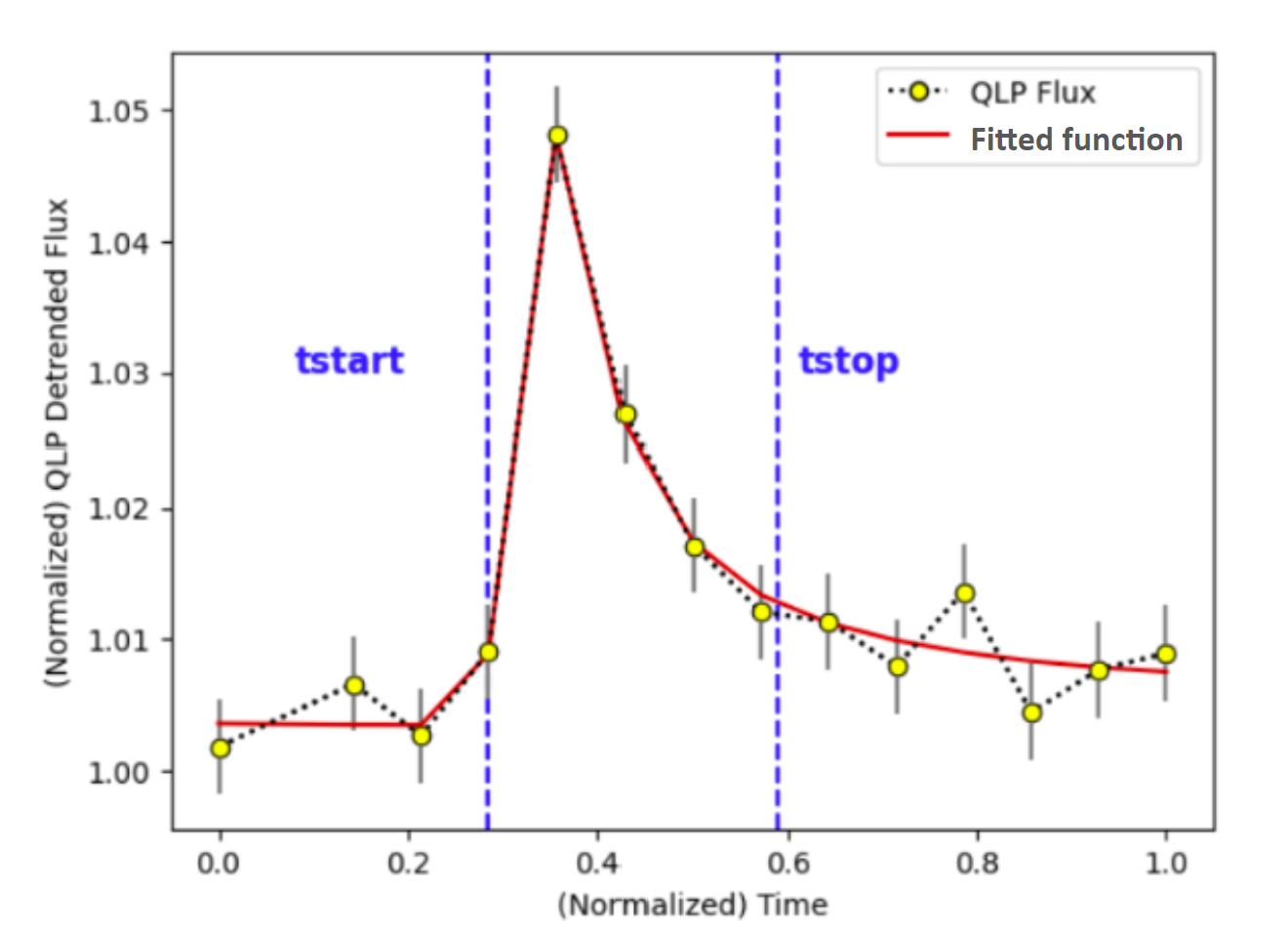}
    \caption{Example of a flare fitted using the "aflare" method, where 'tstop' is defined as 'tstart'+ 'dur' from \cite{Davenport_2016}, and the flare shape function takes the form of the aflare function added to a quadratic background term: $\text{aflare(time,tstart,tdur,amplitude)} + A*\text{time}**2 + B*\text{time} + C$}
    \label{fig:fit}
\end{figure}

The stellar flare detection process involved two major steps: identifying flare candidates and fitting these candidates to derive the relevant physical parameters.

\vspace{2.5mm}

2.3.1 \textit{Flare Candidate Identification}

\vspace{2.5mm}

When a star has more than one TESS sector data, we analyzed each sector LC separately. We  first identified flare candidates in each sector light curve using the 'find\_flares' function from the 'AltaiPony' library \citep{Ilin2021, Davenport_2016}, which implements an algorithm to detect positive deviations in each light curve characteristic of stellar flares, which takes as input three integers $N_{1}$, $N_{2}$, and $N_{3}$. The criteria for flare detection were based on the following conditions \citep{Chang_2015}:

\begin{enumerate}
    \item Positive Excursions: Candidate data points had to be positive excursions from the light curve's quiescent flux.
    \item Sigma Threshold: The excursions had to be $N_{1}\sigma$ above the local scatter of the light curve.
    \item Combined Sigma Threshold: The sum of the positive offset and calculated flux error value had to be $N_{2}\sigma$ above the local scatter.
    \item Consecutive Points: $N_{3}$ or more consecutive data points need to satisfy the first three criteria.
\end{enumerate}

 We experimented with different combinations of $N_{1}$, $N_{2}$, and $N_{3}$ on a small subset (around ~10\% of the data) of the lightcurves and compared the manually confirmed flare events to determine the acceptable parameters. We set $N_{1}$, $N_{2}$, and $N_{3}$ to be $3$, $2$, and $1$, respectively, to allow for appropriate candidate flare detection. This selection allowed for leniency in candidate selection due to factors such as the varying cadences in the TESS data, different signal-to-noise ratios, or different variation types in the continuum fluxes, while still imposing a limit that would mitigate the number of false positive candidates. Some flares were missing from the automatic flare detection and recovery process, which may have been due to our selection of $N_{1}$, $N_{2}$, and $N_{3}$. However, if we relaxed these values by lowering each of them to '1', then it would create a number of candidate flare events (in this particular set of light curves, around 50,000 to 60,000) where only 50\% of them are real.

\vspace{2.5mm}

2.3.2 \textit{Flare Fitting}

\vspace{2.5mm}

Altaipony's flare candidate selection provides flare onset time (tstart), flare duration (t\_dur), and flare amplitude for each flare candidate. We used these values as the initial guess and fit a function to precisely estimate flare parameters. The flare shape function mimics the behavior of stellar flares, a sharp rise in brightness followed by a slow, long decay, and we tried three different functional shapes: a Levy-Stable function, a Pulse function (Blaj et al. 2017), and Altaipony's aflare function (as shown in Figure \ref{fig:fit} with a quadratic background term), which yielded similar fitted parameters. We decided to use Altaipony's aflare function with a quadratic background term for each flare candidate to fit tstart, tdur, amplitude, and three coefficients for the background term. Recovered flares were defined as those with a fitted flare amplitude greater than $3\sigma_{std}$ above the quiescent flux of the light curve, where $\sigma_{std}$ represents the standard deviation after performing sigma clipping on the light curve to remove outliers and reduce the influence of extreme values. The sigma clipping process iteratively excludes data points deviating by more than a set threshold (typically also $3\sigma$) from the median flux, refining the calculation of the background noise and ensuring a more robust detection of true flare events. If the candidate flare amplitude was greater than $0.01$, we sigma-clipped the entire light curve; otherwise we sigma-clipped the zoomed-in light curve region that focused on the flare.  We zoomed in on the light curve region surrounding the candidate flare by selecting the time window \textit{istart-2} to \textit{istart+8}. This window was used for both 30 minute and 10 minute cadence flares.

This approach allowed us to zoom in sufficiently around the flare without including excessive surrounding data. Additionally, the zoomed region was carefully checked for proximity to the edges of the light curve, adjusting the window to ensure that no indices exceeded the bounds of the data. 


\vspace{5mm}

2.3.3 \textit{Conversion to physical units}

\vspace{5mm}

Based on various methods in the literature, we considered a few possible approaches for converting the detected flares from raw TESS flux (in units of electron counts per second) to physical energy units. First, we converted each fitted-flare amplitude to a fractional flare energy value, by expressing the flare energy in terms of the quiescent stellar flux level ($E_{0}$).

\begin{equation}
    {
    {E_{\text{flare}} / E_{0} = f_{\text{flare }} \equiv \displaystyle{\frac{A_{\text{flare}}}{A_{\text{q}}}}}
    }
\end{equation}

where $A_{flare}$ is the area under the curve of the flare in the light curve (calculated by numerically integrating the light curve), and $A_{\text{q}}$ is the area under the quiescent level of stellar flux.

The first method made use of the following relation from \cite{Ealy2024}, where we first calculate the flux $F_{\text{TESS}}$ in the TESS band, then use that value to get the energy of the flare:

\begin{align}
    F_{\text{TESS}} = F_{0} \times 10^{-m_{\text{T}}/2.5} \\
    ED = E_{\text{flare}} / E_{\text{0}} \times (t_{\text{dur,rec}}) \times 86400\\
    E_{\text{flare}} = \text{ED} \times 4\pi d^{2} \times F_{\text{TESS}} 
\end{align}

where $F_{0}$ is zero-point flux in the TESS band ($F_{0} = 4.03 \times 10^{-6} \: \text{ergs} \: \text{cm}^{-2} \: \text{s}^{-1}$ as derived in \cite{Sullivan_2015}), $m_{T}$ is the TESS band apparent magnitude, ED is the equivalent duration of the flare, $t_{\text{dur,rec}}$ is the recovered flare duration in days, and $d$ is the Gaia distance in cm.

The second method we considered first made use of the stellar evolution model from \citep{Baraffe+2015} to estimate the bolometric luminosity of the star based on its effective temperature. Then, using the TESS bolometric correction values from \cite{Eker_2023} we computed the TESS band flux correction for each object, which was multiplied by the bolometric luminosity of the star, and then the equivalent duration of the flare, to the get the estimated flare energy.

The third method uses the TESSreduce package produced by \cite{Ridden_Harper2021}, to calibrate from TESS counts to physical flux. TESSreduce performs background subtraction, field star calibration, and applies an AB magnitude zeropoint to convert the extracted light curves into flux units (e.g., erg s$^{-1}$ cm$^{-2}$), providing a consistent flux calibration across TESS sectors by leveraging reference stars within the cutout field. Then, using relevant distance information as well as the width of the TESS bandpass, we calculated the energy of each flare.

Each of the three methods validated and confirmed each other, giving appropriate flare energy ranges as expected based on previous studies. We hereafter use the flare energy calculation from \cite{Ridden_Harper2021} to produce the relevant figures that involve flare energy, $E_{\text{flare}}$.


\subsection{Injection and recovery of synthetic flares}
To evaluate the flare detection methodology, we performed a synthetic flare injection/recovery test. To do this, we created 30 synthetic flares in each of the QLP lightcurves considered in this study. We created synthetic flares between $\pm 10 \%$ away from the beginning and ending times of the lightcurve and separated from any of the previously identified flare candidates by the average flare duration of that cadence. They were generated with random onset times, an amplitude range of $10^{-3}$ to $10^{2}$ (in units of normalized flux), and a duration range of $0.0001$ to $0.01$ days.

\begin{figure*}
    \centering
    \includegraphics[scale=0.5]{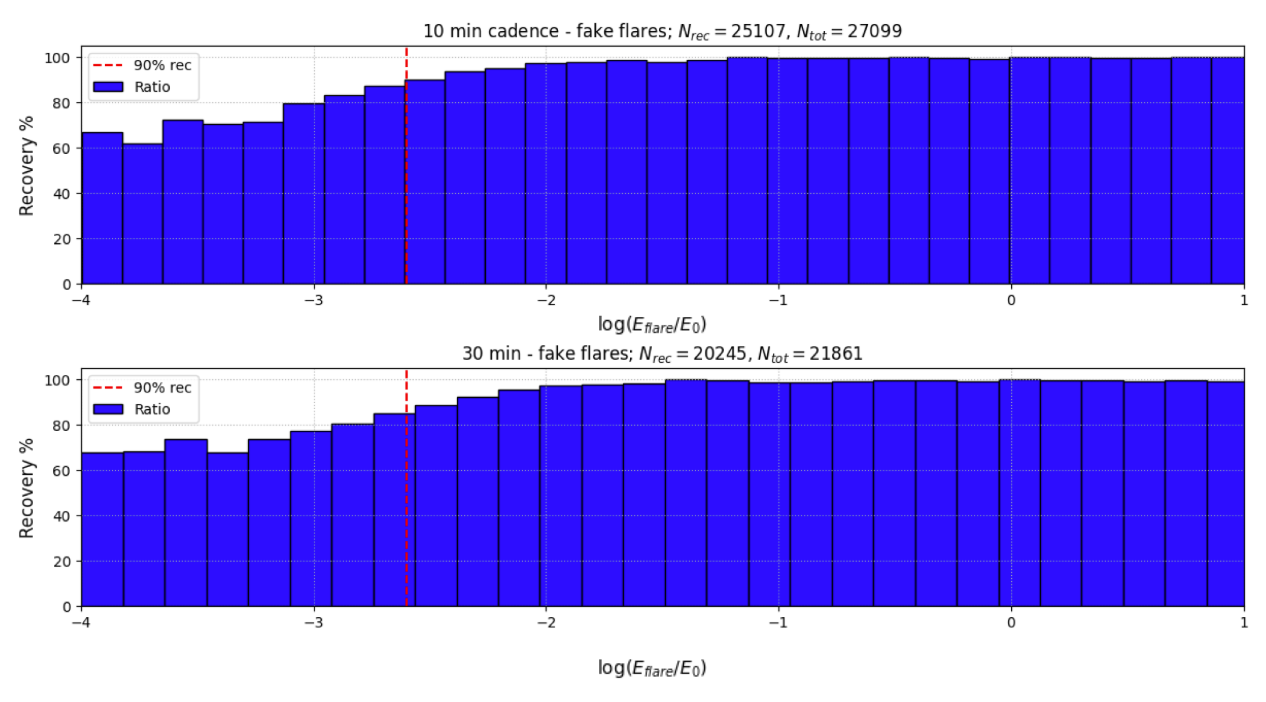}
    \caption{Recovery efficiency of fake flares as a function of flare energy ratio ($E_{\text{flare}}/E_{\text{0}}$) for both 10-minute (top) and 30-minute (bottom) cadence data. The blue bars represent the percentage of fake flares recovered at each energy level, with the total number of recovered flares $N_{\text{rec}}$ and total number of injected flares $N_{\text{tot}}$ indicated for each cadence set. The red dashed line marks the energy level at which the recovery efficiency reaches 90\%. For both cadences, the recovery efficiency increases with flare energy, achieving over 90\% recovery for flares with $E_{\text{flare}}/E_{\text{0}}$ greater than approximately $10^{-2.8}$. The higher cadence (10-minute) data shows a slightly lower energy threshold for achieving 90\% recovery, highlighting the enhanced sensitivity of the 10-minute cadence data in detecting fainter flares.}

    \label{fig:rec_percent}
\end{figure*}

The fake flare experiment, depicted in Figure \ref{fig:rec_percent}, covered a fractional flare energy range from $10^{-4}$ to $10^{1}$. A total of 48,960 fake flares (27099 in the 30 minute cadence data and 21861 in the 10-minute cadence data) were injected into the light curves. We recovered more than 90 \% of them, with $E_{\text{flare}}/E_{\text{0}} \geq 10^{-3}$. These results validate the flare detection algorithms and help set the detection limits.

The analysis of fake flare recovery efficiency reveals key insights into the performance of the flare detection algorithm across different observational cadences. For both the 10-minute and 30-minute cadence data, the algorithm achieves a 90\% recovery rate at a $\text{log}_{10}(E_{\text{flare}}/E_{\text{0}})$
  value of approximately -2.8, and a 70\% recovery rate at approximately -3.3. Below this energy threshold, the recovery rate declines further, indicating a reduction in the algorithm's ability to detect lower-energy flares effectively. The consistent increase in recovery rate with energy across both cadences further confirms the algorithm's effectiveness in identifying flares of varying intensities.

While the discrepancy in recovery rates between the 10-minute and 30-minute cadence data is near negligible, the 10-minute cadence exhibits a slightly lower 90\% recovery threshold. This slight difference underscores the enhanced detection sensitivity of the 10-minute cadence, which allows for more precise and detailed flare detection. The higher time resolution in the 10-minute cadence data play a role in capturing finer details of flare activity, particularly for lower-energy flares that might be missed in the 30-minute cadence data.

\begin{table*}
\centering
\begin{tabular}{lcccccc}
\hline
Moving Group & Age (Myr) & $N_{\text{det}}$ & $N_{\text{poor,removed}}$ & $N_{\text{edge,removed}}$ & $N_{\text{noisy,removed}}$ & $N_{\text{rec}}$ \\
\hline
TW Hydrae (TWA) & 10 & 687 & 268 & 17 & 3 & 153 \\
32 Orionis (Thor) & 22 & 821 & 279 & 41 & 2 & 299 \\
Beta Pictoris & 24 & 2862 & 921 & 205 & 94 & 594 \\
Columba & 42 & 5037 & 1288 & 134 & 39 & 1144 \\
Tucana-Horologium & 45 & 7469 & 1982 & 333 & 141 & 2206 \\
Argus & 50 & 1795 & 836 & 32 & 13 & 133 \\
Carina & 60 & 2336 & 1016 & 61 & 7 & 487 \\
Volans-Carina & 90 & 781 & 427 & 13 & 9 & 109 \\
AB Doradus & 150 & 5628 & 1683 & 229 & 761 & 1163 \\
\hline
Total &  & 27416 & 8700 & 1065 & 1069 & 6288 \\
\hline
\end{tabular}
\caption{Summary of detected flare candidates, removed flares at each cleaning step, and final recovered flares for each moving group.}
\label{table:flares_summary}
\end{table*}

\subsection{Flare Cleaning}

Following the flare detection and recovery process, we implemented several cleaning procedures to remove false flare events which could have resulted from issues such as poor data quality, flares appearing near the edge of light curves, or rotational spikes within faster rotating stars being falsely classified as flare events. This was a multi-step process:

\begin{enumerate}
    
    \item \textbf{Step 1: Flares landing on poor data quality}: we automatically removed any flares that landed on light curve data points that had a non-zero 'QUALITY' flag in the relevant FITs file column, to avoid contamination from instrumental artifacts or data gaps. In total, we removed 8700 flare candidates in this step.
    \item \textbf{Step 2: Flares near the edge of the light curve}: For each flare in our sample, we first determined the associated light curve and located the flare’s onset time (tstart) within it. We then defined a buffer zone with a width three times the 70th percentile of all flare durations, and discarded any flare for which the time between tstart and either the start or end of the light curve was less than this buffer. This removed flares for which the full profile may not be captured due to edge effects. In total, we removed 1065 flare candidates in this step.
    \item \textbf{Step 3: Flares from noisier light curves}: For particularly noisy light curves, we applied an additional filtering step to reject false positives due to residual systematics or high-amplitude background variation. We isolated a small time window around the candidate flare onset and extracted only those data points with 'QUALITY' $= 0$. To quantify the significance of each flare, we computed the global noise level of the light curve using sigma-clipped statistics on the detrended flux values. We then compared the recovered flare amplitude against this global noise, rejecting any candidate whose amplitude did not exceed 3 standard deviations. To avoid bias from flux spikes close to the flare, we excluded a small region around the flare peak and computed the distribution of local peak heights. In this step, we removed 1069 flare candidates.
\end{enumerate}

After the above steps, we checked whether each flare candidate was at least $3\sigma$ above the quiescent flux. If the candidate met this condition, we counted it as a 'recovered' flare,  Additionally, all subsequent analysis omit flares with a $E_{\text{flare}} / E_{\text{0}}$ value smaller than $10^{-3}$, based on the results of the synthetic flares experiment. Following the above processes, we were left with 6288 stellar flares from 1272 light curves of 417 NYMG members, as shown in the last column of Table \ref{table:flares_summary}.

\section{Results} \label{sec:floats}

\subsection{General Flare Statistics}


\begin{table*}
\centering
\setlength{\tabcolsep}{2pt} 
\begin{tabular}{lcccccccccc}
\hline
Member ID & Sector & Spectral Type & Moving Group & Age (Myr) &
\(t_{\text{start,rec}}\) (days) &
\(t_{\text{dur,rec}}\) (days) &
ED (sec) &
\texttt{rec\_ampl} &
\(E_{\text{flare}} / E_{0}\) &
\(E_{\text{flare}}\) (ergs) \\
\hline
HIP25436 & 33 & F4IV/V & Columba & 42 & 2221.5959 & 0.015625 & 1.45 & 0.00500 & 0.0011 & 4.53$\times10^{33}$ \\
TYC\_8595-1740-1 & 10 & G5 & Carina & 60 & 1575.0757 & 0.012360 & 1.81 & 0.02350 & 0.0017 & 4.56$\times10^{32}$ \\
TYC\_8595-1740-1 & 10 & G5 & Carina & 60 & 1590.5964 & 0.082394 & 30.64 & 0.01338 & 0.0043 & 7.70$\times10^{33}$ \\
TYC\_8595-1740-1 & 37 & G5 & Carina & 60 & 2312.6378 & 0.007689 & 2.58 & 0.04068 & 0.0039 & 6.50$\times10^{32}$ \\
TYC\_8595-1740-1 & 35 & G5 & Carina & 60 & 2264.1791 & 0.005641 & 2.26 & 0.05121 & 0.0046 & 5.67$\times10^{32}$ \\
TYC\_8595-1740-1 & 36 & G5 & Carina & 60 & 2287.6448 & 0.004821 & 2.91 & 0.10752 & 0.0070 & 7.32$\times10^{32}$ \\
TYC\_8595-1740-1 & 36 & G5 & Carina & 60 & 2298.5198 & 0.006342 & 1.00 & 0.01793 & 0.0018 & 2.52$\times10^{32}$ \\
TYC\_8595-1740-1 & 9 & G5 & Carina & 60 & 1549.6172 & 0.008825 & 1.39 & 0.04169 & 0.0018 & 3.49$\times10^{32}$ \\
TYC\_8595-1740-1 & 9 & G5 & Carina & 60 & 1550.1589 & 0.027631 & 3.42 & 0.01073 & 0.0014 & 8.60$\times10^{32}$ \\
TYC\_8595-1740-1 & 9 & G5 & Carina & 60 & 1551.0131 & 0.009484 & 1.22 & 0.01889 & 0.0015 & 3.07$\times10^{32}$ \\
TYC\_8595-1740-1 & 9 & G5 & Carina & 60 & 1552.5131 & 0.046875 & 11.41 & 0.01257 & 0.0028 & 2.87$\times10^{33}$ \\
TYC\_8595-1740-1 & 9 & G5 & Carina & 60 & 1552.8464 & 0.008784 & 0.82 & 0.01437 & 0.0011 & 2.06$\times10^{32}$ \\
TYC\_8574-2094-1 & 9 & K0 & Carina & 60 & 1547.9294 & 0.038662 & 111.14 & 0.21192 & 0.033 & 2.77$\times10^{33}$ \\
TYC\_8574-2094-1 & 9 & K0 & Carina & 60 & 1559.2835 & 0.083225 & 101.13 & 0.04385 & 0.014 & 2.52$\times10^{33}$ \\
TYC\_8574-2094-1 & 9 & K0 & Carina & 60 & 1561.7001 & 0.110794 & 178.71 & 0.04910 & 0.019 & 4.46$\times10^{33}$ \\
TYC\_8574-2094-1 & 9 & K0 & Carina & 60 & 1562.5959 & 0.005736 & 1.99 & 0.11483 & 0.0040 & 4.97$\times10^{31}$ \\
TYC\_8574-2094-1 & 9 & K0 & Carina & 60 & 1563.3668 & 0.025215 & 79.06 & 0.35371 & 0.036 & 1.97$\times10^{33}$ \\
TYC\_8574-2094-1 & 10 & K0 & Carina & 60 & 1571.6375 & 0.033639 & 32.07 & 0.05622 & 0.011 & 8.00$\times10^{32}$ \\
TYC\_8574-2094-1 & 10 & K0 & Carina & 60 & 1573.1167 & 0.029788 & 26.61 & 0.05582 & 0.010 & 6.63$\times10^{32}$ \\
TYC\_8574-2094-1 & 10 & K0 & Carina & 60 & 1573.5125 & 0.054072 & 40.50 & 0.03944 & 0.0087 & 1.01$\times10^{33}$ \\
TYC\_8574-2094-1 & 10 & K0 & Carina & 60 & 1573.9292 & 0.009556 & 34.05 & 0.97810 & 0.041 & 8.49$\times10^{32}$ \\
TYC\_8574-2094-1 & 10 & K0 & Carina & 60 & 1574.9917 & 0.031044 & 15.57 & 0.03073 & 0.0058 & 3.88$\times10^{32}$ \\
TYC\_8574-2094-1 & 10 & K0 & Carina & 60 & 1575.1792 & 0.002885 & 1.01 & 0.03633 & 0.0041 & 2.52$\times10^{31}$ \\
TYC\_8574-2094-1 & 10 & K0 & Carina & 60 & 1575.4083 & 0.000650 & 0.23 & 0.04301 & 0.0042 & 5.85$\times10^{30}$ \\
TYC\_8574-2094-1 & 10 & K0 & Carina & 60 & 1577.0749 & 0.028143 & 51.35 & 0.19813 & 0.021 & 1.28$\times10^{33}$ \\
TYC\_8574-2094-1 & 10 & K0 & Carina & 60 & 1580.8666 & 0.008136 & 7.17 & 0.22137 & 0.010 & 1.79$\times10^{32}$ \\
TYC\_8574-2094-1 & 10 & K0 & Carina & 60 & 1581.6166 & 0.005678 & 24.85 & 1.31195 & 0.051 & 6.20$\times10^{32}$ \\
TYC\_8574-2094-1 & 10 & K0 & Carina & 60 & 1584.9290 & 0.008853 & 10.57 & 0.33566 & 0.014 & 2.64$\times10^{32}$ \\
TYC\_8574-2094-1 & 10 & K0 & Carina & 60 & 1587.2623 & 0.050705 & 83.25 & 0.08787 & 0.019 & 2.08$\times10^{33}$ \\
TYC\_8574-2094-1 & 10 & K0 & Carina & 60 & 1589.6164 & 0.010113 & 5.85 & 0.17008 & 0.0067 & 1.46$\times10^{32}$ \\
\hline
\end{tabular}
\caption{Detected flare events for stars in nearby young moving groups (NYMGs).
The columns represent: the member (star) identifier, TESS sector number, spectral type, moving group,
$t_{\text{start,rec}}$ (recovered flare start time in BTJD days),
$t_{\text{dur,rec}}$ (recovered flare duration in days),
equivalent duration (ED) in seconds,
\texttt{rec\_ampl} (recovered flare amplitude),
$E_{\text{flare}} / E_{0}$ (fractional flare energy),
and flare energy in ergs.
Values are rounded to astrophysically meaningful precision.
The full table of all 6288 flares is available in machine-readable form in the online journal.}
\label{table:flares_table}
\end{table*}


 In this section, we analyze the cumulative flare frequency distributions (FFDs) for the 9 NYMGs in Table \ref{table:flares_summary}, as functions of stellar age and mass, for the 6288 flares that we recovered. A subset of these flares is tabulated in Table \ref{table:flares_table}.

For the FFD calculations and relevant figures that consider flare frequency, we had to consider an appropriate way to calculate the rate of flaring. We used a per-star flare rate value for these plots, first by calculating how many flares were detected for a particular star, which we called $N_{\text{flare}}$. We then calculate the total effective duration of observation ($T_{\text{days}}$) using only non-zero quality flag data points, then took $N_{\text{flare}}/ T_{\text{days}}$ to get the flare rate for that star.

\begin{figure*}
    \centering
    \includegraphics[scale=0.7]{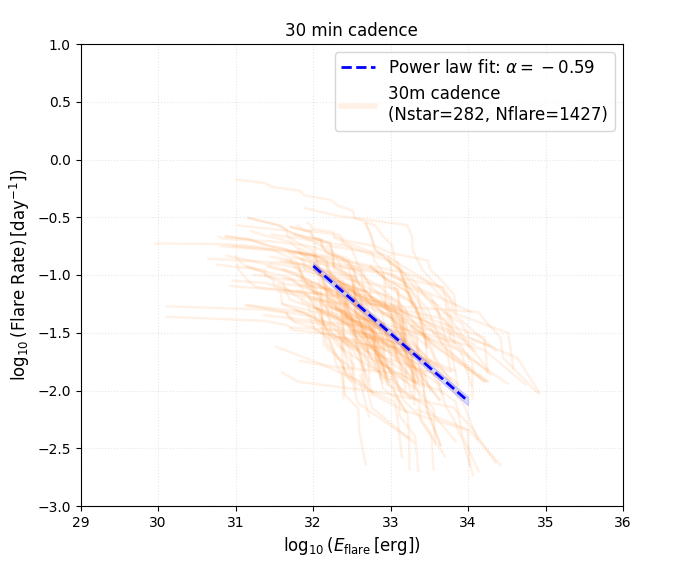}
    \includegraphics[scale=0.7]{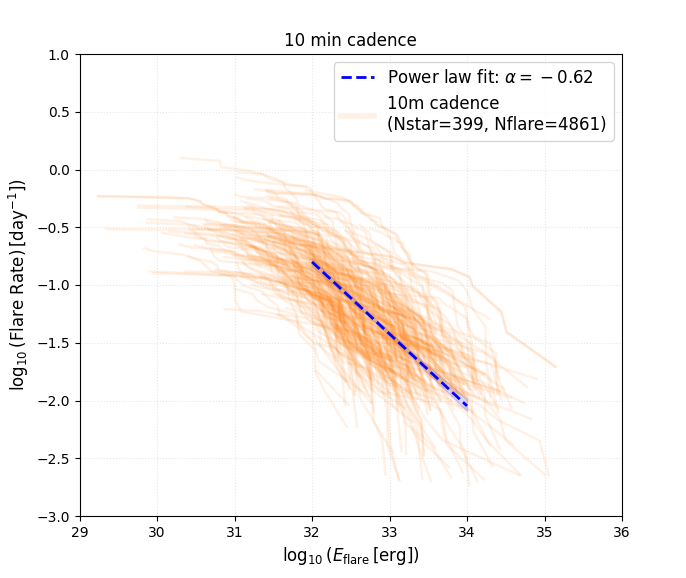}
    \caption{FFDs for flares detected in the 30-minute cadence data (top) and 10-minute cadence data (bottom). These panels include both automatically and manually detected flares. The x-axis shows the log flare energy $E_{\text{flare}}$ in ergs, and the y-axis shows the flare rate detected at each energy level. The number of flares detected, $N_{\text{flare}}$, is provided in the legend for each cadence.}

    \label{fig:real_ffd}
\end{figure*}

The analysis of flares in units of ergs, shown in Figure \ref{fig:real_ffd}, reveals a significant difference in detection counts: 1427 flares were recovered in the 30-minute cadence data, compared to 4831 flares in the 10-minute cadence data. The smaller sample of 10-minute cadence light curves (479 LCs as opposed to 788 for 30-minute) highlights the increased sensitivity and resolution afforded by higher cadence observations, which is crucial for understanding stellar magnetic activity and its impact on stellar and planetary environments. Both FFDs show a trend consistent with the expected power law distribution of stellar flare activity.

\begin{figure*}
    \centering
    \includegraphics[width=1.0\linewidth]{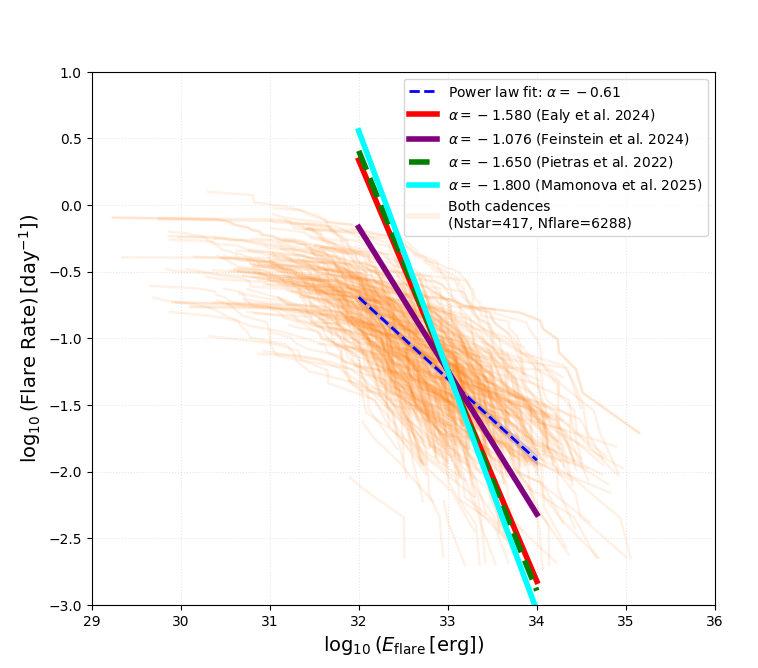}
    
    \caption{The flare frequency distributions (in ergs) for the entire data, with a power law fit from this work and \cite{Ealy2024}, \cite{feinstein_2024}, and \cite{Pietras2022}}
    \label{fig:ffd_comparison}
\end{figure*}

\begin{figure}
    \centering
    \includegraphics[scale=1.0]{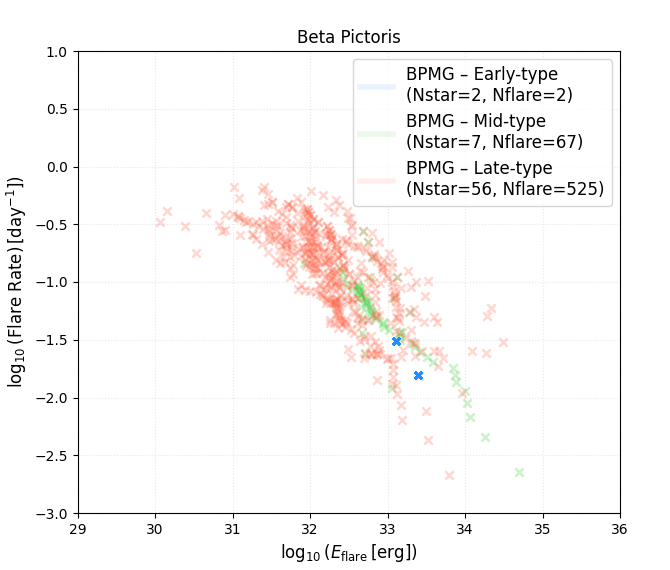}
    \caption{Flare frequency distribution (FFD) for the Beta Pictoris Moving Group (BPMG, 24 Myr), divided by Bp-Rp color index.}

    \label{fig:BPMG_ffd}
\end{figure}

\begin{figure}
    \centering
    \includegraphics[scale=1.0]{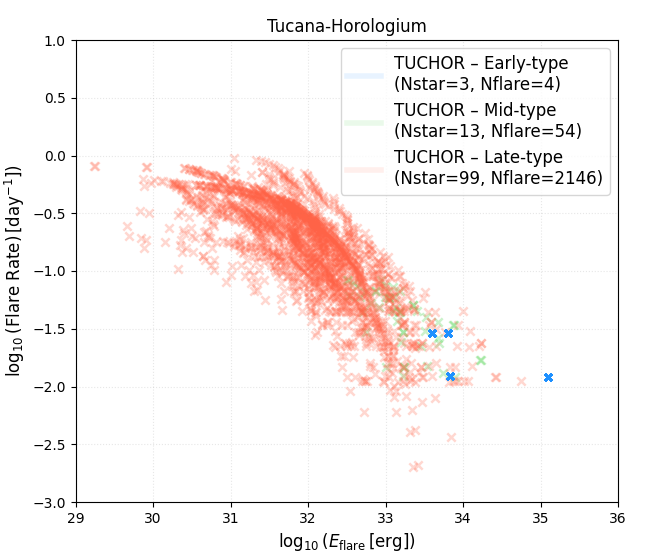}
    \caption{Flare frequency distribution (FFD) for the Tucana-Horologium Moving Group (TucHor, 45 Myr).}

    \label{fig:tuchor_ffd}
\end{figure}

\begin{figure}
    \centering
    \includegraphics[scale=1.0]{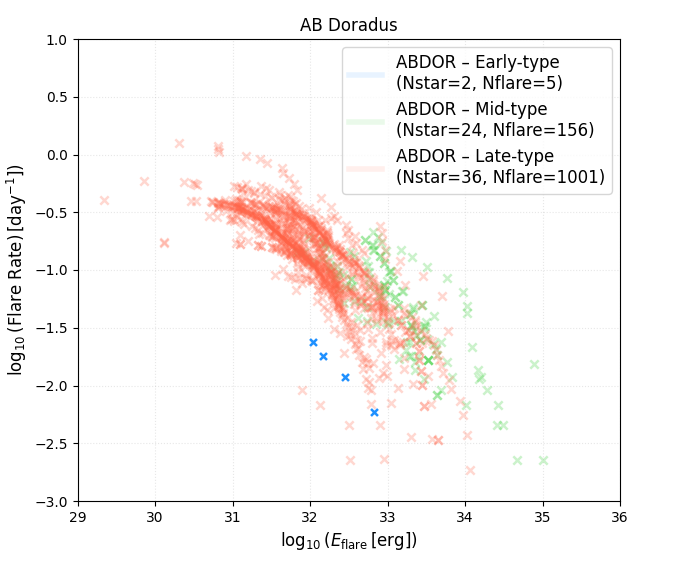}
    \caption{Flare frequency distribution (FFD) for the AB Doradus Moving Group (ABDor, 150 Myr). Flare frequency distributions for 6 additional moving groups (TucHor, ABDor, Columba, etc.) will be available in the online figure set.}

    \label{fig:abdor_ffd}
\end{figure}

If stellar flares are produced through a slow-driven energy dissipation system with a stationary energy input rate and a fixed threshold for triggering a flare (i.e., SOC model; \cite{Aschwanden_2019}), we can expect a power-law relationship between flare energies and frequencies. This expectation can be extended to a set of homogeneous stars with similar age, mass, rotation period, and magnetic cycle phase (i.e., a similar magnetic activity environment). However, our NYMG sample is a heterogeneous group of stars that may not share such a same magnetic activity environment. In the distributions of flares (Figures \ref{fig:real_ffd} - \ref{fig:abdor_ffd}), we see a broader range of distributions, likely due to the mixture of flaring stars with different characteristics. Yet, we can still see the average power law behavior (of $\approx-0.6$ to $-0.5$) from the ensemble of individual stars' FFDs.


In Figure \ref{fig:ffd_comparison}, we show the complete ensemble of individual FFDs for the full sample of detected flares in comparison to previous flare studies. We applied a Bayesian hierarchical linear regression to the ensemble of FFDs to derive a global power-law slope of the densest region of the plot. The model fits $\text{log}_{10}\nu = \alpha \text{log}_{10}\text{E}+\beta$, where $\nu$ is the flare rate, and E is the calculated flare energy, over the range $\text{log}_{10}\text{E}\in [32,34]$. Sampling with PyMC yields posterior estimates of the global slope and its credible interval, shown as the blue line in Figure \ref{fig:ffd_comparison}. Outside this interval, the distribution departs from a single power law because of incompleteness at low E and small-number noise at high E. At around $10^{31}$, this marks the approximate completeness boundary of our sample; below this energy, the rates should be regarded as lower limits. The fitted power law for this distribution differs from those reported in previous flare studies. \cite{Pietras2022}, who analyzed flares from a larger sample of 330,000 stars from TESS, found a steeper power law index of $-1.65$ (in the energy range of $5 \times 10^{33}$ to $10^{36}$ ergs), while \cite{Ealy2024} reported an index of $-1.58$ for a study of BPMG and AB Doradus (in the energy range of $10^{31}$ to $10^{35}$ ergs). Our findings for the complete NYMG sample yield shallower slopes than those two studies, but are closest to the power law slope found by \cite{feinstein_2024}, who found a slope of $-1.076$ for stars with Rossby numbers lower than $0.136$. \cite{Mamonova_2025}, who also studied several NYMGs in this age range, found a significantly steeper power law slope of $-1.80$ in the roughly $10^{32}$ to $10^{34}$ energy range for AU Microscopii, an active M-dwarf. Although this is one star, its slope provides a useful point of comparison for assessing whether flare statistics in NYMGs at the population level show similar underlying scaling behavior. Additionally, at higher flare energies, the ensemble of FFDs broadly trends downward, reflecting possible variations in energy band sensitivity. The discrepancies in the power law fits could arise from the independent SOC nature of the various differing characteristics (age, mass, etc.) between each study's individual sample of stars. These differences could also come from the varying rotation period measurements in each sample (such as the selection of fast rotators in \cite{feinstein_2024}), which can influence the magnetic dynamo/topology and thus the extent and overall behavior of flaring energy.

In Figures \ref{fig:BPMG_ffd} - \ref{fig:abdor_ffd}, we divided the stars into three mass and spectral type bins. The high-mass bin corresponds to stars with a Gaia Bp-Rp color index less than or equal to $0.8$, which includes F type stars. The intermediate-mass bin encompasses stars with a Bp-Rp color index between $0.8$ and $1.6$, covering G-type stars and early K-type stars. Finally, the low-mass bin includes stars with a Bp-Rp color index greater than or equal to $1.6$, which corresponds to late K-type and M-type stars. These bins are represented by blue, green, and red curves, respectively, in the FFD analysis plots. 

As shown sequentially in Figure \ref{fig:BPMG_ffd} (BPMG), Figure \ref{fig:tuchor_ffd} (Tucana-Horologium), and Figure \ref{fig:abdor_ffd} (AB Doradus), there is a lack of evidence of changing flare activity behavior with respect to age. In each of these figures, the flare frequency distribution (FFD) is broken down by the Bp-Rp color index; the blue (hot) bin is sparesely populated in all three groups. For each moving group, the members maintain significant flare activity across a broad energy range. Across all three moving group FFDs, cooler, late-type stars dominate the flare population, with a higher amount of flares overall for each distribution, as expected from magnetic activity scaling. There is a subtle indication of changing flare activity for solar-type stars (green points); however, this is at the mercy of small number statistics.

The Tucana-Horologium (TucHor) sample shows a broader dispersion at higher flare energies ($\text{log E} > 33$), likely due to a larger sample of flares. This dispersion could reflect star-to-star variations within the group—differences in rotation rates, magnetic activity levels, or stellar structure—resulting in slightly different scaling behaviors between individual stars. Detection biases could also contribute, as late-type stars with lower photospheric fluxes tend to reveal weaker flares more easily.

However, in some cases, like that of AB Doradus, the scatter of flares from solar-type stars (green) is slightly higher on average. This could suggest the presence of fast rotators in solar-type stars persisting, and could also necessitate the investigation into any binaries within the sample that could lead to binary interaction effects that could also cause spin-up in stars. These factors could cause stronger magnetic activity in G-K stars. This is in line with the findings of \cite{Mamonova_2025}, who proposed rotation period as a means of gauging flare activity, as opposed to age. Nonetheless, it is consistent with previous findings that young, rapidly rotating solar-type stars are capable of producing rare, very energetic superflares \citep{Maehara2012}.

\begin{figure*}
    \centering
    \includegraphics[width=0.75\linewidth]{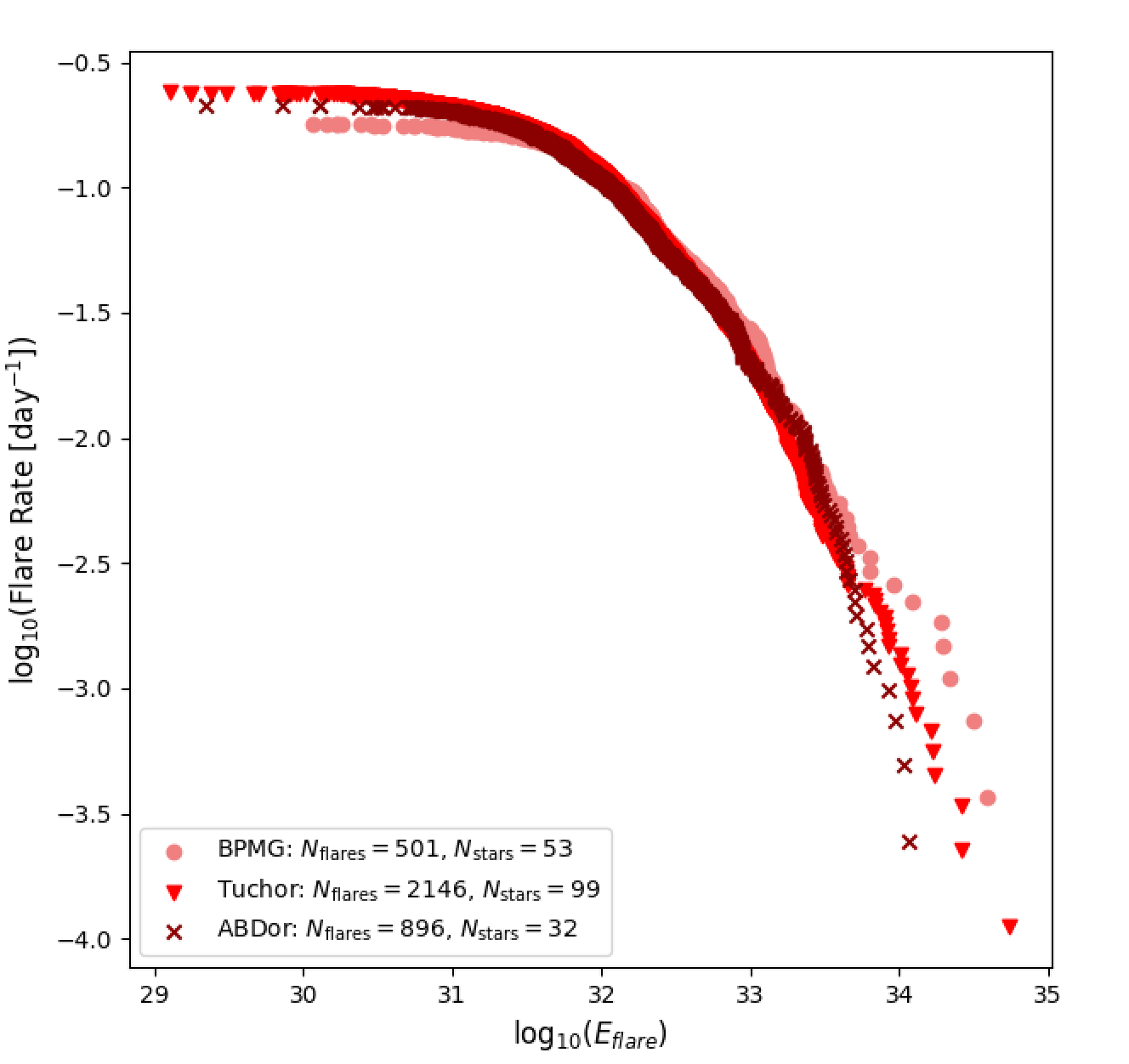}
    \caption{FFD of late type stars with respect to age for BPMG, Tuc-Hor, and AB Doradus. We assume SOC-scaling holds for this sub-sample.}
    \label{fig:FFD_late}
\end{figure*}

Taking a broader look at the flare activity across each moving group, we see subtle trends across later-type stars, as shown in the FFDs in Figure \ref{fig:FFD_late}. Since this plot shows FFD curves that individually come from the same mass bin and moving group, we can assume an SOC-scaling, and thus plot the cumulative FFD curves for each sub-population, opposed to the ensemble of FFD curves for each star, as in the earlier plots. The flare rate generally stays the same as we  progress in moving group age from Beta Pictoris (~24 Myr) to AB Doradus (~150 Myr), but at the higher energy end, the expected decrease in flaring with age is evident.

The FFDs of other moving groups are available as a figure set.

\figsetstart
\figsetnum{10}
\figsettitle{Flare Frequency Distributions for 6 Moving Groups}

\figsetgrpstart
\figsetgrpnum{10.1}
\figsetgrptitle{TW Hydrae
}
\figsetplot{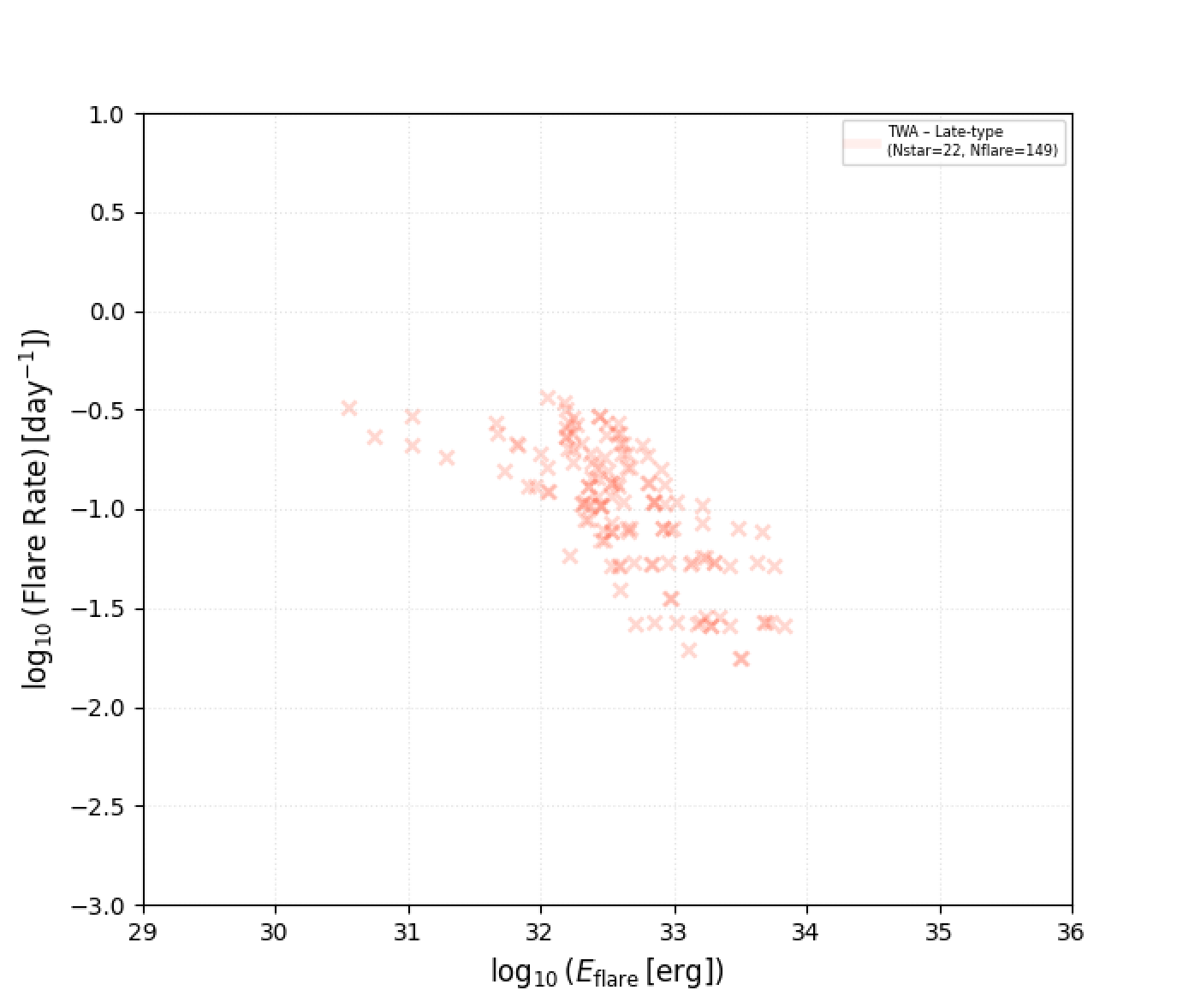}
\figsetgrpnote{Flare frequency distribution for the TW Hydrae Moving Group.}
\figsetgrpend

\figsetgrpstart
\figsetgrpnum{10.2}
\figsetgrptitle{32 Orionis (Thor)
}
\figsetplot{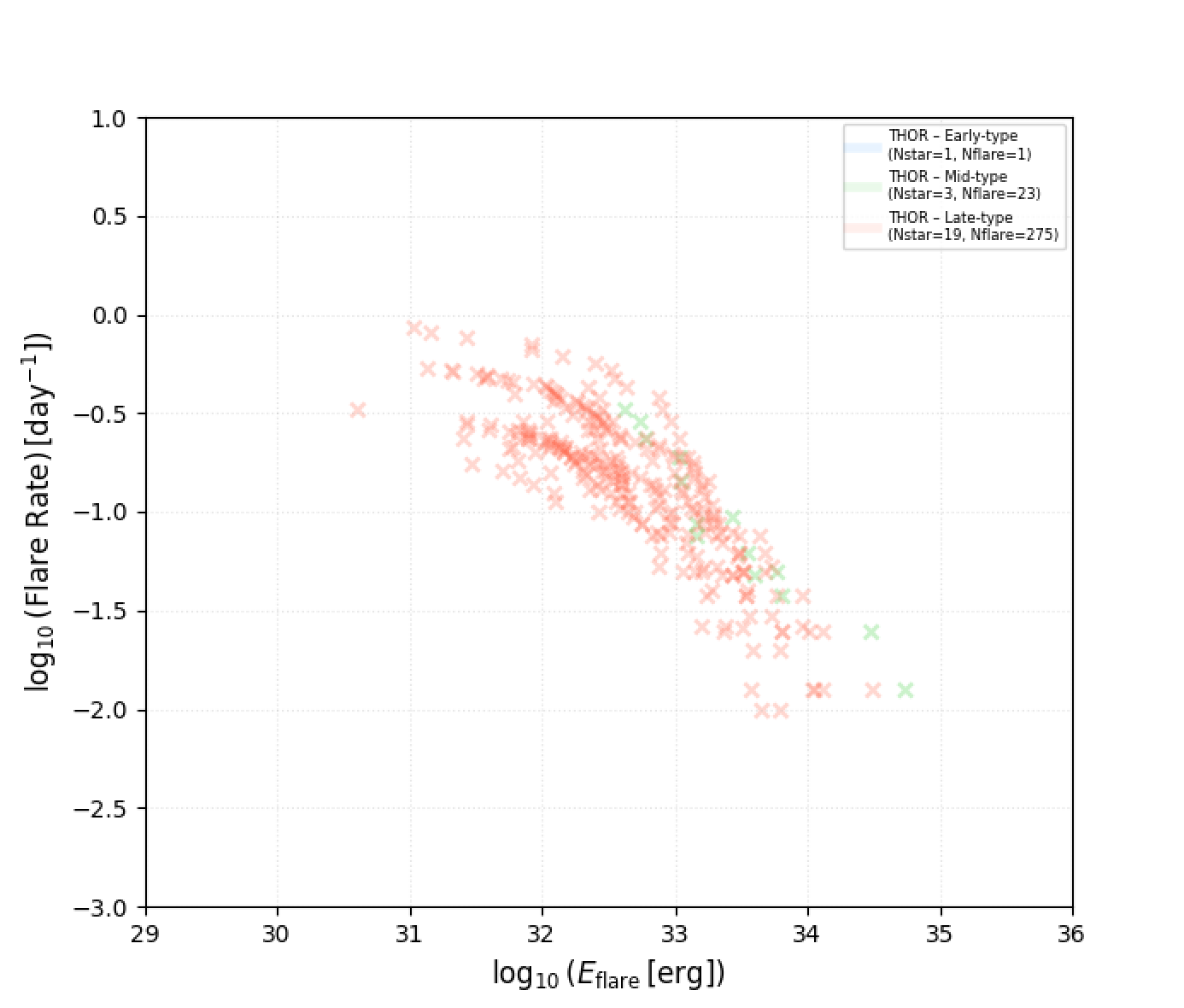}
\figsetgrpnote{
Flare frequency distribution for the 32 Orionis Moving Group.
}
\figsetgrpend

\figsetgrpstart
\figsetgrpnum{10.3}
\figsetgrptitle{Columba
}
\figsetplot{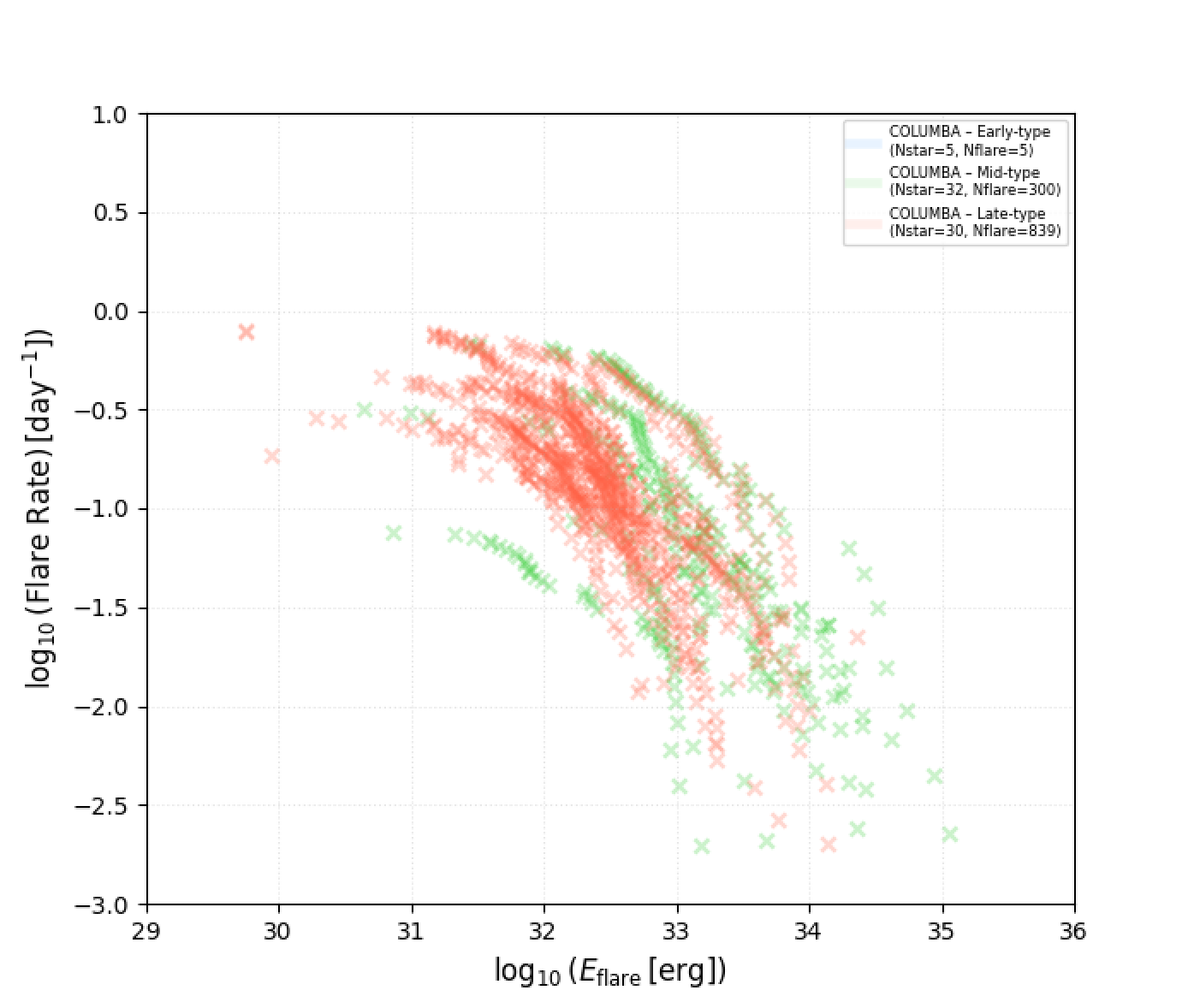}
\figsetgrpnote{
Flare frequency distribution for the Columba Moving Group.
}
\figsetgrpend

\figsetgrpstart
\figsetgrpnum{10.4}
\figsetgrptitle{Argus
}
\figsetplot{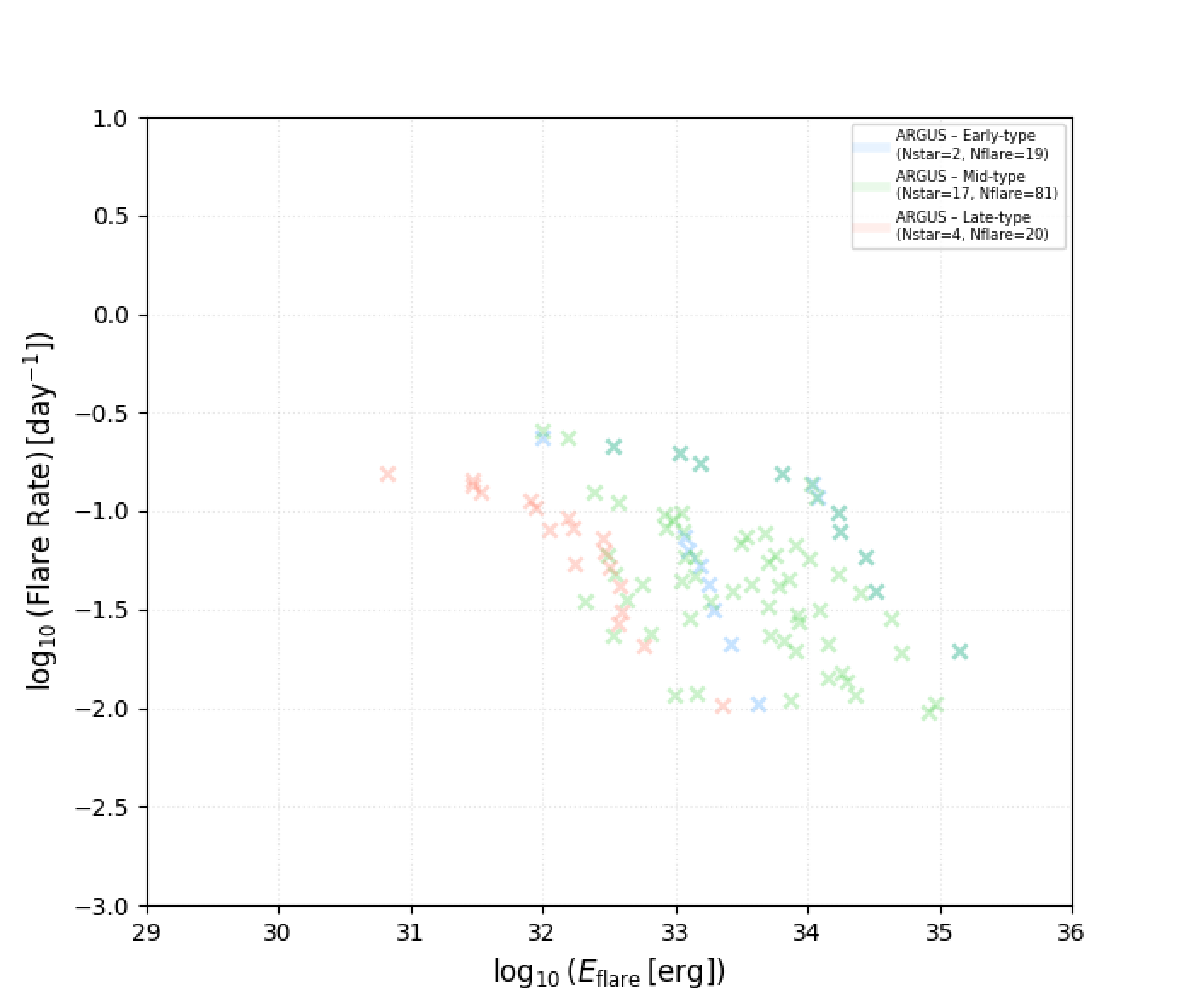}
\figsetgrpnote{
Flare frequency distribution for the Argus Moving Group.
}
\figsetgrpend

\figsetgrpstart
\figsetgrpnum{10.5}
\figsetgrptitle{Carina
}
\figsetplot{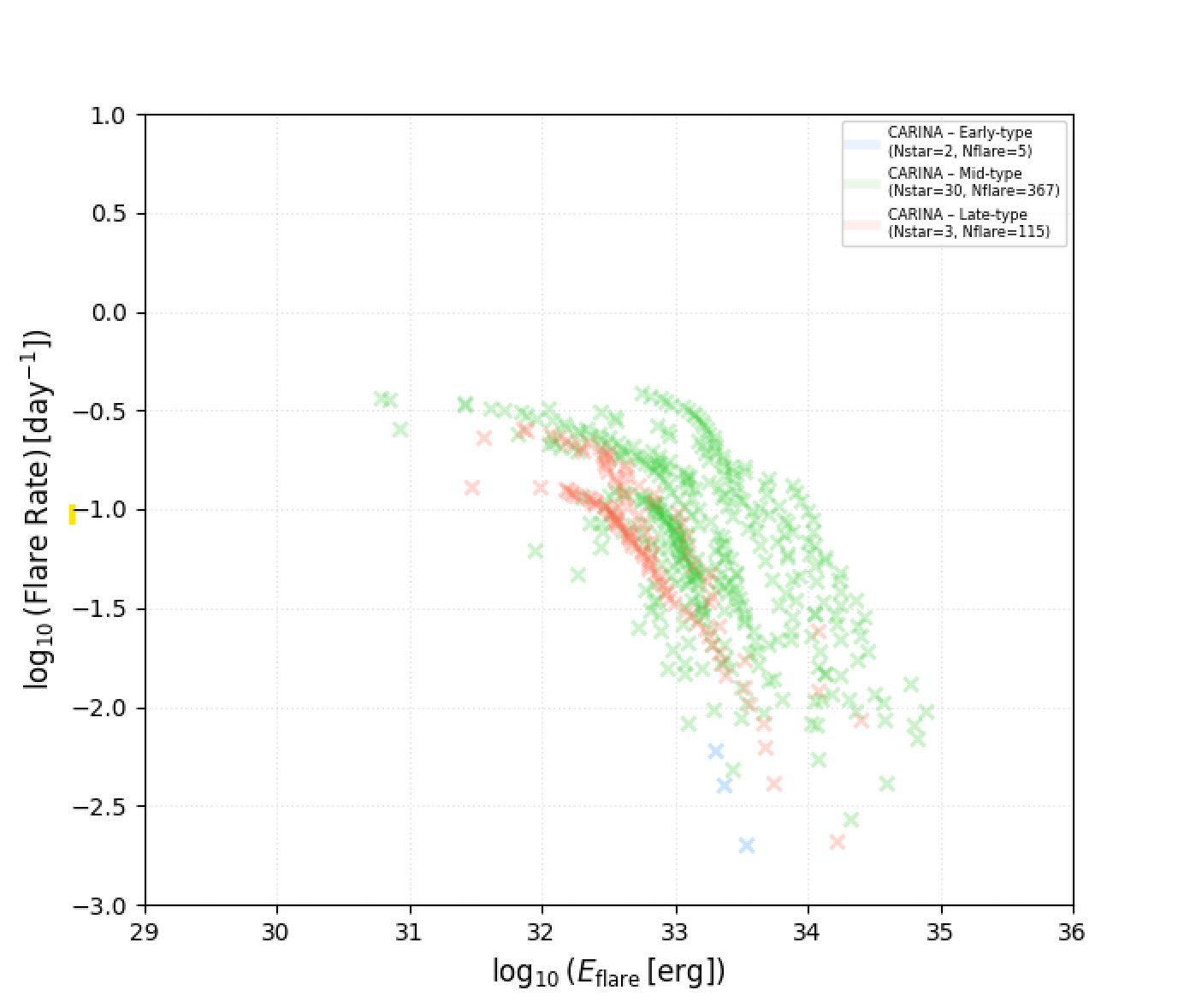}
\figsetgrpnote{
Flare frequency distribution for the Carina Moving Group.
}
\figsetgrpend

\figsetgrpstart
\figsetgrpnum{10.6}
\figsetgrptitle{Volans-Carina}
\figsetplot{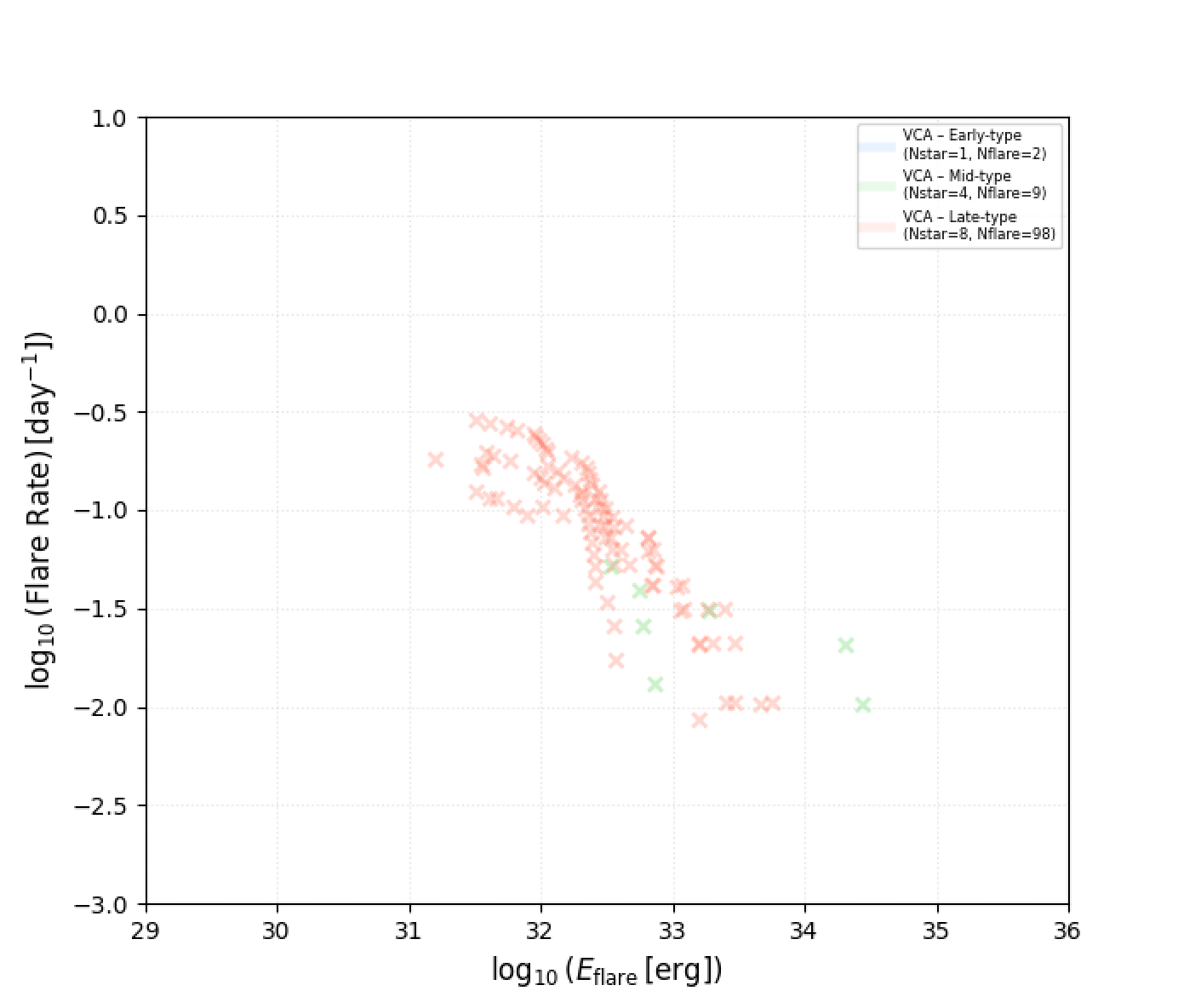}
\figsetgrpnote{
Flare frequency distribution for the Volans-Carina Moving Group.}
\figsetgrpend

\figsetend

\begin{figure*}
    \centering
    \includegraphics[scale=0.75]{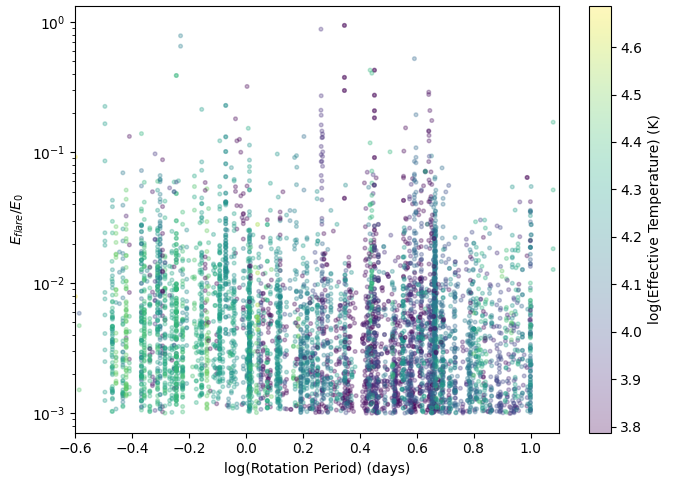}
    \caption{Fractional flare energy ($E_{\text{flare}}/E_{\text{0}}$) plotted with respect to rotation period and temperature.}
    \label{fig:rot_temp}
\end{figure*}

\begin{figure*}
    \centering
    \includegraphics[scale=0.575]{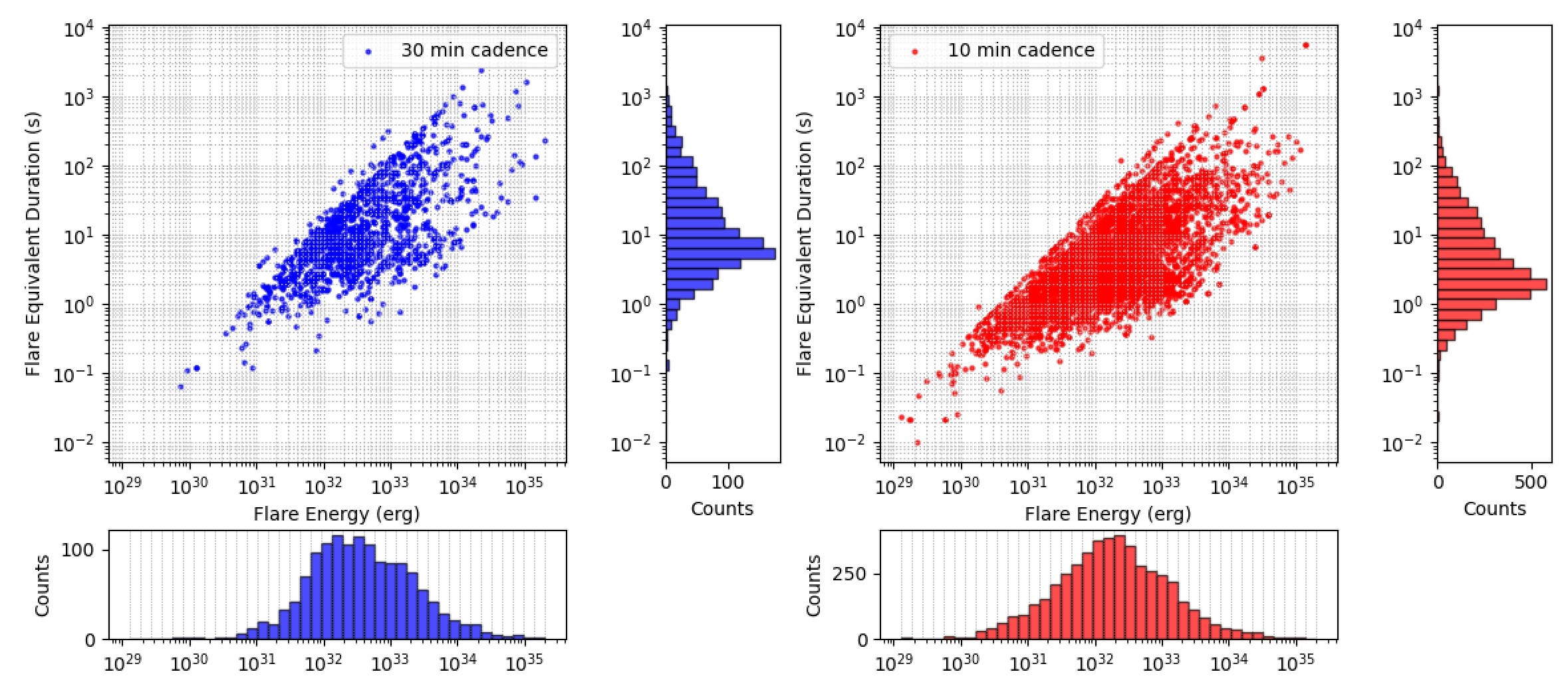}
    \caption{Scatter plot showing the relationship between flare energy (in erg) and flare duration (in days) for 30-minute cadence data (left, blue) and 10-minute cadence data (right, red). The data reveal a positive correlation between flare energy and duration, consistent with models of stellar flare activity. The higher cadence (10-minute) data shows a broader range of detected flare energies and durations, including slightly more short-duration flares, indicating enhanced sensitivity to both low-energy short-duration and high-energy long-duration flares.}

    \label{fig:dur}
\end{figure*}

\begin{figure*}
    \centering
    
    \includegraphics[scale=1.5]{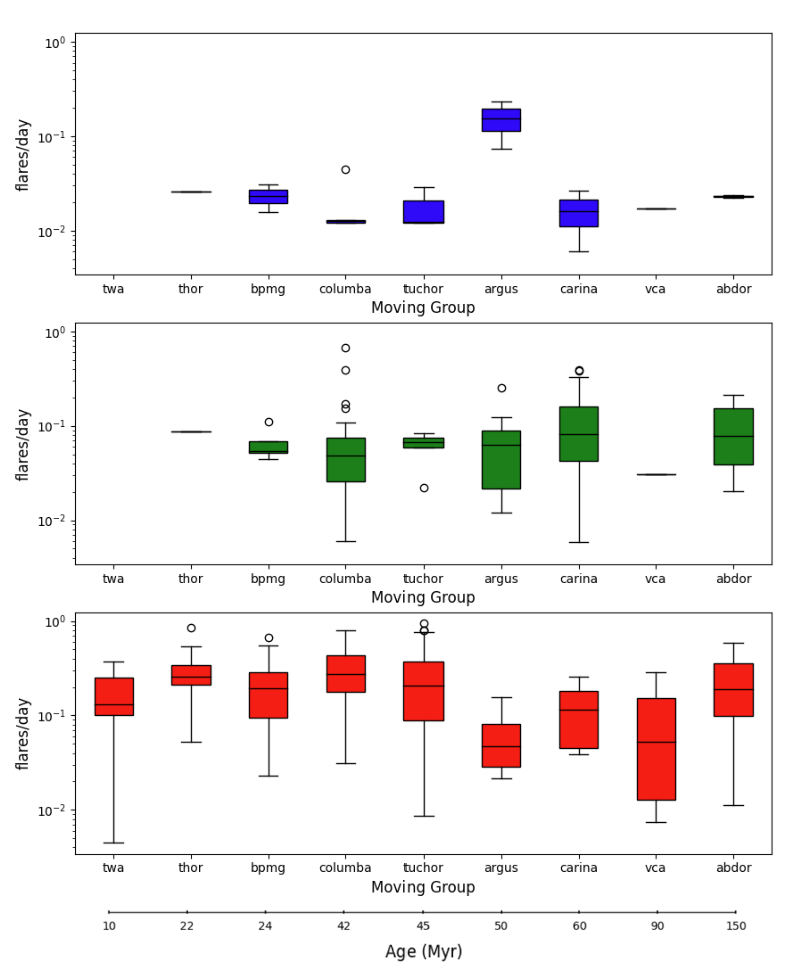}

    \caption{Average flare rate (number of flares per day per star) for members of different young moving groups (MGs), categorized by spectral type. The panels display average flare rates for stars of different Gaia BP-RP color indices, indicating their spectral types: \textit{Top: } stars with a Gaia BP-RP color index less than or equal to 0.75, covering A and F spectral types. The total number of stars in this bin is 18. \textit{Middle: } stars with a Gaia BP-RP color index between 0.75 and 1.8, corresponding to G-type stars and early K-type stars. The total number of stars in this bin is 89. \textit{Bottom}: stars with a Gaia BP-RP color index greater than or equal to 1.8, representing late K-type and M-type stars. The late-type stars show an average flare rate roughly 5-6 times higher than early type stars. The total number of stars in this bin is 310.}

    \label{fig:rate}
\end{figure*}

\begin{figure}
    \centering
    \includegraphics[width=1.0\linewidth]{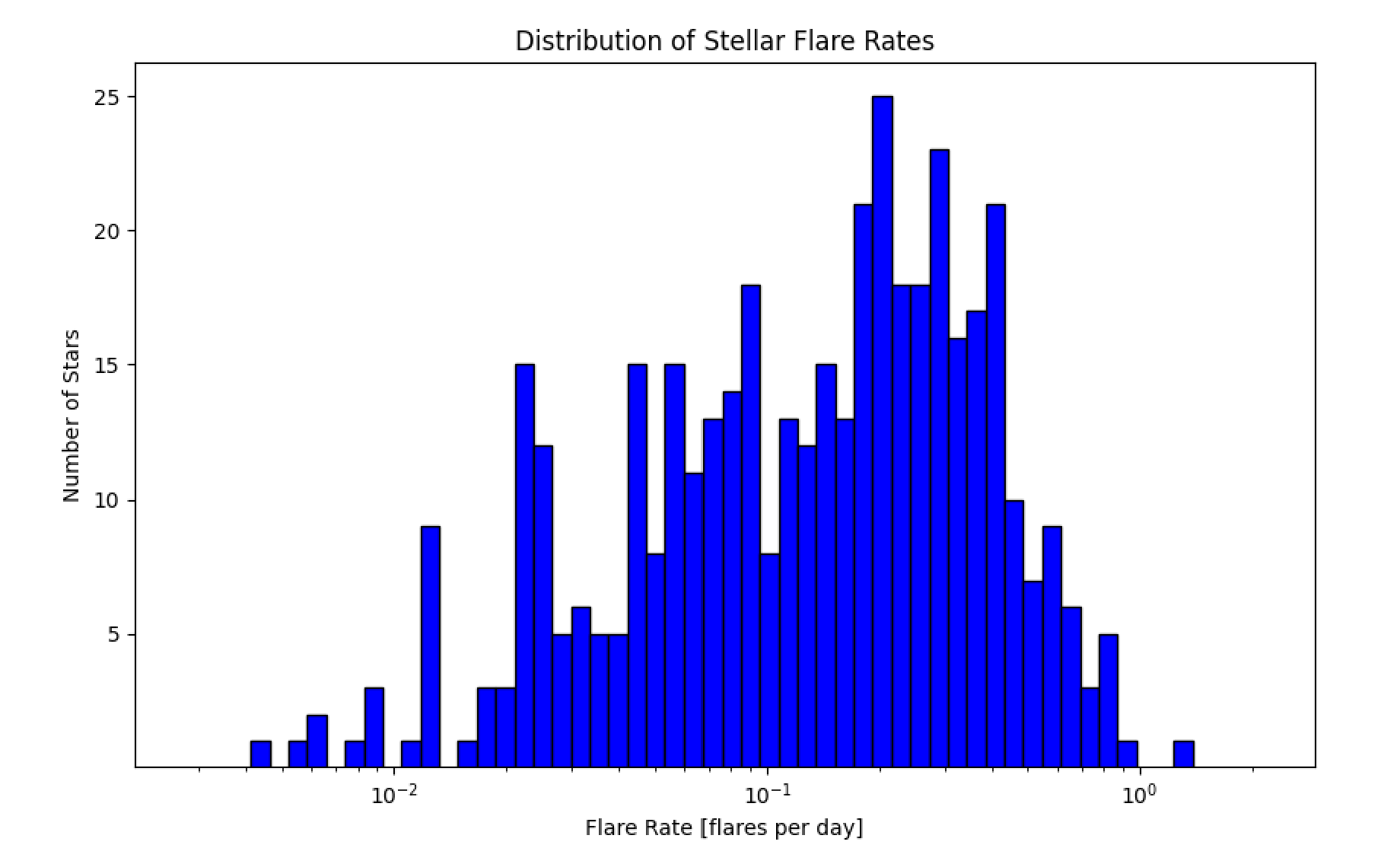}
    \caption{Distribution of flare rate (in units of flares/day) for the flaring stars in the NYMG sample.}
    \label{fig:flare_rate}
\end{figure}


\vspace{5mm}


In Figure \ref{fig:rot_temp}, we show the relationship between fractional flare energy, stellar rotation period, and effective temperature. There is some noticeable vertical structure, possibly due to the binning of the rotation period values (i.e. plotting many flares that correspond to a star with a certain rotation period value). However, the distribution of flare energy appears scattered across the rotation period axis, indicating no strong correlation between flare energy and rotation period in this sample. While there is a hint of a concentration of cooler, later-type stars around a log $($Rotation Period$)$ $0.2$ to $0.6$ days, the scatter in the data suggests that the relationship between flare energy and rotation is complex and likely influenced by additional stellar properties not accounted for in this analysis, or that flare activity is saturated over this age range.

The flare events span a wide energy range from approximately $10^{29}$ erg to $10^{36}$ erg and equivalent durations from $10^{-2}$ seconds to $10^{4}$ seconds, as shown in Figure \ref{fig:dur}. Notably, the 10-minute cadence data detect flares with slightly shorter minimum equivalent durations compared to the 30-minute cadence. This increased sensitivity and higher resolution of the 10-minute cadence is expected, as more frequent observations enhance the capability of detecting of short-equivalent duration flares. The highest energy flares recovered in this study reach the superflare threshold as defined by \cite{Maehara2012}. The strong correlation between flare energy and equivalent duration in both cadence regimes is expected, as equivalent duration is directly derived from the flare energy calculation. The observed spread in the distribution likely reflects differences in flare morphology, particularly the varying rates of flux decay in individual light curves, which influence how extended the flare appears in time.



Figure \ref{fig:rate} shows the distribution of the average flare rate for each moving group, once again split into 3 spectral-type bins. The main distribution of flare rates are mostly within a range of just over 0.01 to 1 flares per day, as indicated by the distribution shown in Figure \ref{fig:flare_rate}. This distribution of flare rates is consistent with previous studies like \cite{feinstein_2024} who equally analyzed young stars but up to a higher age of 300 Myr. Interestingly, there is no clear trend indicating a strong age dependence,  but, as expected, cooler, redder stars exhibit higher flare rates compared to their hotter, bluer counterparts, which could be related to the convective zones and overall magnetic activity within cooler stars. However, the Argus moving group shows a higher level of flare activity in its early type members compare to its later type members which have lower flare activity, which necessitates a revaluation of Argus in terms of age and membership. It is worth noting that in the blue bin, the flare activity is coming from a total of 18 stars across all groups, which may only partly explain the lower observed flare rates in the blue bin in general (and not just due to the general lack of flare activity in early type stars).

The lack of a connection between effective temperature and flare energy implies that other factors are likely more influential in determining the energy of stellar flares. Factors such as magnetic field strength, magnetic flux distribution, and the star's magnetic cycle might play more crucial roles in driving flare activity. This observation is consistent with the understanding that magnetic activity, rather than temperature alone, governs the occurrence and intensity of flares. For instance, most A-type stars do not possess magnetic fields, with the exception of a few that have static fields, but do not have the capability to produce stellar flares \citep{2004Natur.431..819B}. This spectral class was relatively underrepresented in our sample, with only 26 out of 590 unique stars being A-type. This ensures that the majority of the stars we analyzed likely have magnetic fields capable of driving flares.

\subsection{Higher-Energy Flare Events}

We find that there are several flare events that land in the energy range of $10^{33}$ ergs to $10^{35}$ ergs from solar and late type stars. Previous studies have provided various threshold definitions of "superflares"; the seminal work of \cite{Maehara2012} used a threshold of $10^{33}$ ergs (for the analysis of solar type stars), while \cite{Vasilyev+2024} used $10^{34}$ ergs as their threshold, and found that superflares occur only once per century on sun-like stars.

Since we are working with a broader range of spectral types and younger stars, we choose the threshold of $10^{34}$ ergs, in line with \cite{Vasilyev+2024}, in order to place an emphasis on the most extreme extraordinary "megaflare" type events. This is after accounting for stellar luminosity and spectral type scaling relations from \cite{Zombeck2010}, which highlight that stellar luminosity varies by several orders of magnitude from F to M type stars, making luminosity ratio $L/L_{\odot}$ a crucial scaling factor when converting flare measurements such as equivalent duration into flare energies.

Table \ref{table:flares_table2} shows a sample of the flare events above this designated threshold. Figure \ref{fig:FFD_late} shows that at such high energies, there is an indication that older moving groups flare less frequently than younger ones. While intermediate flare energies are more or less consistent across all three moving groups, the highest energy flares show a clear evolution. At lower energies beyond the detection limit of $\text{log E} \approx 31$, the difference cannot be justified, and has to be taken cautiously. 

If we concentrate on the tail end of each distribution, we see that in the megaflaring energy range, the median flare rate is around $10^{-3.5}$, or roughly $0.0003$ flares per day. Extrapolating from this value, we can make the assumption that flares of this caliber occur around 11 times per century on late-type stars (about 10 times higher than older, sun-like stars; see \cite{Vasilyev+2024}). For M-dwarf hosts, repeated flaring at this scale may erode planetary atmospheres \citep{do_Amaral_2025}, while for solar-type stars, rare but extreme superflares could intermittently alter the radiation environment of surrounding planets.

\begin{table*}
\centering
\setlength{\tabcolsep}{1pt} 
\begin{tabular}{lcccccccccc}
\hline
Member ID & Sector & Spectral Type & Moving Group & Age (Myr) &
\(t_{\text{start,rec}}\) (days) &
\(t_{\text{dur,rec}}\) (days) &
ED (sec) &
\texttt{rec\_ampl} &
\(E_{\text{flare}} / E_{0}\) &
\(E_{\text{flare}}\) (ergs) \\
\hline
PMM\_4809 & 36 & G3e & Argus & 50 & 2285.3185 & 0.02910 & 38.35 & 0.0491 & 0.015 & 3.79$\times10^{34}$ \\
HD\_12894 & 3 & F2V & Tuc--Hor & 45 & 1387.2847 & 0.06507 & 34.52 & 0.0221 & 0.0061 & 1.37$\times10^{35}$ \\
TYC\_8595-1740-1 & 37 & G5 & Carina & 60 & 2318.6933 & 0.05625 & 77.00 & 0.0328 & 0.016 & 1.94$\times10^{34}$ \\
TYC\_8595-1740-1 & 37 & G5 & Carina & 60 & 2321.5544 & 0.05625 & 41.17 & 0.0175 & 0.0085 & 1.03$\times10^{34}$ \\
TYC\_8595-1740-1 & 37 & G5 & Carina & 60 & 2324.2002 & 0.05625 & 51.03 & 0.0217 & 0.011 & 1.28$\times10^{34}$ \\
TYC\_8595-1740-1 & 35 & G5 & Carina & 60 & 2263.4986 & 0.03774 & 162.92 & 0.1334 & 0.050 & 4.10$\times10^{34}$ \\
TYC\_8595-1740-1 & 36 & G5 & Carina & 60 & 2287.0128 & 0.05625 & 60.43 & 0.0257 & 0.012 & 1.52$\times10^{34}$ \\
TYC\_8595-1740-1 & 36 & G5 & Carina & 60 & 2296.2906 & 0.05563 & 66.14 & 0.0285 & 0.014 & 1.66$\times10^{34}$ \\
TYC\_8595-1740-1 & 36 & G5 & Carina & 60 & 2299.5684 & 0.03262 & 65.79 & 0.0691 & 0.023 & 1.65$\times10^{34}$ \\
TYC\_8595-1740-1 & 36 & G5 & Carina & 60 & 2303.8809 & 0.01421 & 42.92 & 0.1887 & 0.035 & 1.08$\times10^{34}$ \\
TYC\_8595-1740-1 & 9 & G5 & Carina & 60 & 1550.6589 & 0.10543 & 42.11 & 0.0122 & 0.0046 & 1.06$\times10^{34}$ \\
2MASS\_J09345645-64595 & 37 & F5V & VCA & 90 & 2317.5683 & 0.01610 & 6.64 & 0.0248 & 0.0048 & 2.37$\times10^{34}$ \\
HD\_302321 & 10 & F8 & Carina & 60 & 1590.9509 & 0.02136 & 6.61 & 0.0400 & 0.0036 & 1.08$\times10^{34}$ \\
HD\_85151B & 9 & G9V & Argus & 50 & 1564.3060 & 0.12744 & 708.23 & 0.1551 & 0.064 & 1.70$\times10^{34}$ \\
HD\_85151B & 36 & G9V & Argus & 50 & 2288.1460 & 0.04317 & 687.36 & 0.4497 & 0.18 & 1.65$\times10^{34}$ \\
HD\_85151B & 36 & G9V & Argus & 50 & 2290.4099 & 0.01990 & 472.73 & 1.0297 & 0.27 & 1.14$\times10^{34}$ \\
HD\_85151B & 36 & G9V & Argus & 50 & 2301.8959 & 0.02983 & 1101.90 & 1.3509 & 0.43 & 2.65$\times10^{34}$ \\
HD\_85151B & 35 & G9V & Argus & 50 & 2263.2499 & 0.05625 & 4582.36 & 1.9512 & 0.94 & 1.10$\times10^{35}$ \\
HD\_85151B & 35 & G9V & Argus & 50 & 2275.0140 & 0.04132 & 1066.33 & 0.7340 & 0.30 & 2.56$\times10^{34}$ \\
\hline
\end{tabular}
\caption{The first 20 flares in the dataset above the designated threshold of $10^{34}$ ergs. The columns represent: the member (star) identifier, TESS sector number, spectral type, moving group,
$t_{\text{start,rec}}$ (recovered flare start time in BTJD days),
$t_{\text{dur,rec}}$ (recovered flare duration in days),
equivalent duration (ED) in seconds,
\texttt{rec\_ampl} (recovered flare amplitude),
$E_{\text{flare}} / E_{0}$ (fractional flare energy),
and flare energy in ergs.
Values are rounded to astrophysically meaningful precision.
The full table of all 183 high energy flare events is available in machine-readable form in the online journal.}
\label{table:flares_table2}
\end{table*}

\section{Summary and Discussion} \label{sec:cite}

In this work, we performed a comprehensive analysis of stellar flares in members of Nearby Young Moving Groups (NYMGs) aged 10-150 Myr using photometric data from the Transiting Exoplanet Survey Satellite (TESS). Our study focused on characterizing the flare frequency distributions (FFDs) and investigating the relationship between stellar flares and physical parameters like age, spectral type, rotation period, and effective temperature. 

Our analysis yielded several important results:

\begin{enumerate}
    \item \textbf{Ubiquity of Flares:}  
    All NYMG stars with available TESS QLP light curves analyzed in this work exhibit one or more stellar flares, confirming that flare activity is nearly universal among young stars within the 10–150 Myr age range.

    \item \textbf{Flare Detection:} Using our flare identification scheme, we could identify 6288 flares from 417 stars at a 90\% detection rate above $E_{\text{flare}}/E_{0}$ values of $10^{-3}$.

    \item \textbf{Age Dependence:}  
    No clear age dependence of flare activity is observed across the $\sim$10–150 Myr range for any spectral type. Over this age range, there may be saturated flare activity, similar to the case of X-ray emission on low-mass stars.

    \item \textbf{Spectral-Type Dependence:}  
    Lower-mass stars (M-type) display flare frequencies approximately 5–6 times higher than those of G–K type stars, and roughly an order of magnitude higher than F-type stars. This behavior is consistent with enhanced magnetic activity in low mass stars.

    \item \textbf{Flare Rates:}  
    NYMG members show average flare rates exceeding 0.01 flares per day over the energy range of $10^{31}$–$10^{33}$ erg. More energetic events ($E > 10^{34}$ erg) occur less frequently, roughly once every few years.

    \item \textbf{Cadence Dependence:}  
    The 10-minute cadence data yield approximately 5.5 times more flares than the 30-minute cadence data over the same energy range. This increase is attributed to the improved temporal sampling of flare peaks at higher cadence. Extrapolating from this result, we expect the 200-second TESS cadence mode to achieve an additional factor of $\sim$2–3 higher flare recovery efficiency.

    \item \textbf{Superflare Fraction:}  
    Based on the adopted threshold of superflares from \cite{Vasilyev+2024} ($E \geq 10^{34}$ erg), approximately 25\% of NYMG members display superflare activity. When the superflare threshold is scaled to the quiescent flux level as a function of spectral type, about 10\% of NYMG members still exhibit superflares.

    \item \textbf{Flare Frequency Distribution (FFD) Slope:}  
    The power-law index of the FFDs, derived from a fit over the $10^{32}$–$10^{34}$ erg range, is $\alpha = -0.51$, which is shallower than values reported in previous studies ($-1.1 \lesssim \alpha \lesssim -1.7$; e.g., \cite{Pietras2022}; \cite{feinstein_2024}). This difference likely arises from the inclusion of a larger number of low-energy flares, which flatten the observed slope.

    \item \textbf{Rate of Flaring:}  
    While no clear trend in overall flare rates with age is seen, early-type stars exhibit lower mean flare rates ($\sim$0.1 flares/day) compared to mid- and late-type stars ($\sim$0.3–0.6 flares/day). The overall flare rate distribution remains consistent across the NYMGs, suggesting stable magnetic activity during this epoch.

    \item \textbf{Argus Moving Group Anomaly:}  
    In the Argus Moving Group, early-type stars show unusually elevated flare activity levels (by $\gtrsim 3\sigma$ or several times higher than similar stars in other groups), while M-type members display comparatively weaker flare activity. These discrepancies suggest that the Argus group’s age and/or membership assignments should be revisited through detailed spectroscopic and kinematic analyses.
\end{enumerate}

Future work should focus on expanding the sample size to include older stellar populations from open clusters (such as the Hyades cluster or Ursa Major Association), field stars, more intermediate and high-mass stars, as well as exploring the long-term activity cycles of NYMG members, using the abundance of TESS data from the past 7 years. Additionally, further analysis of flare activity in different observational cadences (such as Cycle 5 and later from TESS) will enhance our understanding of magnetic field evolution in young stars, as the larger sample of stars and enhanced temporal resolution will equally allow for the increased number of high energy, superflare events as well as the detection of previously missed lower energy flares, respectively. Finally, the flare statistics we collected for young stars may allow us to use them as a proxy for assessing stellar habitability.
\begin{acknowledgments}
The authors would like to thank the anonymous refrees for the comments that improved the quality of this paper. This research was supported by NASA Grant Number 80NSSC25K7805. We acknowledge the use of data from the Transiting Exoplanet Survey Satellite (TESS) mission \citep{https://doi.org/10.17909/fwdt-2x66}, a NASA Explorer mission led and operated by MIT, and funded by NASA's Science Mission Directorate. We thank the TESS team for making this data publicly available and accessible.
\end{acknowledgments}

%

\vspace{5mm}
\facilities{TESS}


\software{astropy \citep{2013A&A...558A..33A,2018AJ....156..123A},  
          Cloudy \citep{2013RMxAA..49..137F}, 
          Source Extractor \citep{1996A&AS..117..393B}
          , altaipony \citep{Ilin2021, Davenport_2016}, lightkurve \citep{2018ascl.soft12013L}}




\bibliography{sample631_final}{}
\bibliographystyle{aasjournal}



\end{document}